# Using Machine Learning to Create an Early Warning System for Welfare Recipients♠

Dario Sansone[1]

Anna Zhu[2]

This version: May 2021

**Abstract**

Using high-quality nation-wide social security data combined with machine learning tools, we develop predictive models of income support receipt intensities for any payment enrolee in the Australian social security system between 2014 and 2018. We show that off-the-shelf machine learning algorithms can significantly improve predictive accuracy compared to simpler heuristic models or early warning systems currently in use. Specifically, the former predicts the proportion of time individuals are on income support in the subsequent four years with greater accuracy, by a magnitude of at least 22% (14 percentage points increase in the $R^2$), compared to the latter. This gain can be achieved at no extra cost to practitioners since the algorithms use administrative data currently available to caseworkers. Consequently, our machine learning algorithms can improve the detection of long-term income support recipients, which can potentially provide governments with large savings in accrued welfare costs.

♠ We thank Bruce Bradbury, Simon Feeny, David McKenzie, Tim Reddel, and Tim Robinson, as well as the participants of seminars delivered at the Social Policy Research Centre, University of New South Wales and the Australian National University, for their helpful comments. Yin King Fok provided excellent research assistance. Zhu acknowledges the support of the Australian Research Council (ARC) Linkage Project (LP170100472). This paper uses unit record data from the Centrelink administrative records from the Department of Social Services (DSS). The findings and views reported in this paper are those of the authors and should not be attributed to the ARC or DSS. All errors are our own.

[1] University of Exeter and IZA. E-mail: d.sansone@exeter.ac.uk
[2] RMIT University and IZA. Email: anna.zhu@rmit.edu.au

# 1. Introduction

Long-term income support (welfare) receipt is an issue many governments around the world aim to prevent (Hanna 2019, HM Government 2010, Reddel 2018, Scoppetta and Buckenleib 2018, Welfare Working Group 2011). In basic security and/or targeted welfare systems, income support payments are designed to provide a minimum standard of living to households who are unable to meet essential consumptions needs with income from private sources (Korpi and Palme 1998). Thus, individuals who regularly receive income support – over an extended period of time – are most likely to suffer long-term economic disadvantage and social exclusion. Entrenched reliance on income support also imposes significant demands on government budgets, reduce economy-wide market output, and lead to the intergenerational transmission of welfare cultures (Cobb-Clark et al. 2017, Dahl et al. 2014, Dahl and Gielen 2020). The problems associated with entrenched income support reliance are likely to intensify as a result of the COVID-19-induced recession as individuals remain unemployed for extended periods.

To prevent such entrenched reliance, policymakers try to intervene early in the welfare careers of high-risk registrants with labour market activation programs and casework management. A first step in early intervention programs, however, is to identify a target group. In Australia, for example, the government is specifically focused on targeting individuals who are most disadvantaged and thus most likely to stay on welfare for a protracted period of time (Department of Employment, Skills 2020, Department of Social Services 2018). Reflecting this target, we develop predictive models of income support receipt intensities for any payment enrolee in the Australian social security system, using high-quality nation-wide social security data combined with Machine Learning (ML) tools.

A key attraction of an Australian case-study is that it has one of the most targeted welfare systems in the world, which means that when we predict the risk of long-term income support receipt we closely predict the incidence of ongoing poverty and social exclusion (Whiteford 2010). Furthermore, our research is extremely policy relevant and timely since the Australian government is currently trialling innovative profiling tools in the targeting of early intervention programs, with the specific aim of preventing welfare dependency (Reddel, 2018). If proven successful, this early warning system could thus act as a blueprint for other countries on resource allocation solutions.

Currently, identifying these at-risk individuals involves simple profiling tools, valuations of lifetime welfare costs, and/or caseworkers' evaluations. Yet, these screening devices can lead to poor targeting of resources for a number of reasons. First, simple screening tools can poorly predict or identify who is at risk of long-term welfare receipt because this outcome is likely the product and interaction of multiple risk factors, which often cannot be predicted with a slim set of early warning indicators (Bradbury and Zhu 2018). Second, frontline staff may make "cream-skimming" targeting decisions, such as skewing the offering of employment activation programs towards those for whom outcomes are easier to achieve. This is especially prevalent for countries that contract out a high percentage (100% in Australia's case) of employment services to private agencies and where compensation is tied to outcomes (O'Sullivan et al. 2019). Third, current practices to identify high-risk registrants also over-burden caseworkers: for example, in Australia, a caseworker at any one time has an average caseload of more than 100 clients (Davidson 2019). Furthermore, this labour-intensive screening process may subsequently divert resources away from individualised intervention efforts. Identifying high-risk individuals with greater accuracy and at a low-cost is a key motivation for this paper.

The data we use are of high quality because the government relies on these exact data to determine an individual's eligibility to income support payments. Since an individual's payment amount is a direct function of their income, wealth, savings, household structure, and several other socio-economic factors, the information in these data are reconciled with Australian Tax Office records to ensure accuracy. These data are also high frequency, with daily information on income support receipt status, following clients from 2000 to 2019. As these data are rich in covariates (approximately 1,800 possible predictors), large in size (with over 32 million individual clients), and are easily accessible by caseworkers, they are ideal for calibrating ML algorithms and for devising an early warning system for long-term welfare recipients. We identify those people who are most likely to suffer long-term disadvantage and who consequently will receive income support.[3] Our results are thus relevant to the targeting of two types of policy responses – those that address need, and those which seek to reduce government transfers.

---

[3] It is worth noting that welfare receipt is not necessarily a problem in and of itself since it can serve to protect the health and well-being of recipients, particularly those who face unexpected shocks or major structural impediments to work (Aizer and Currie 2004, Mitrut and Tudor 2018). In contrast, entrenched reliance on income support is considered an unintended cost of work disincentives embedded in program designs, such as that of highly targeted welfare systems (Feldstein 2005, Hoynes 1997, Hoynes and Schanzenbach 2012, Moffitt 1985, 1992).

Our focus is on the outcome of any income support, since this is also a focus of the Australian government. Given its emphasise in the past economic literature (Card et al. 2007, 2015, Schmieder and von Wachter 2016), we also estimate as an extension ML algorithms for the outcome of long-term receipt of unemployment-specific benefits. Unemployment benefit recipients are a subset of all income support recipients. A distinguishing feature of benefit eligibility for job-seekers is that they have additional job search and activity requirements to meet, compared to a recipient of other types of income support payment such as disability benefits or parenting allowances.

The topic of our research speaks to two previous bodies of work. The first is the nascent literature on estimating ML models to inform resource allocation decisions. This literature has shown that ML algorithms can assist personnel with day-to-day decisions because they better predicts at-risk individuals compared to standard regression tools (Kleinberg et al. 2015). For example, judges can improve bail-granting decisions (Kleinberg et al. 2017), health inspectors can be more efficiently allocated to unhygienic restaurants (Glaeser et al. 2018, Kang et al. 2013), program administrators can better target interventions at high-risk youth (Chandler et al. 2011), employers can make better hiring decisions (Hoffman et al. 2018), schools can promptly identify students at risk of dropping out (Sansone 2019), government officials can prevent child abuse and maltreatment (Cuccaro-Alamin et al. 2017, Vaithianathan et al. 2013), and surgeons can more ethically and effectively screen patients for hip-replacement surgery (Mullainathan and Spiess 2017). In addition, researchers have combined ML with satellite data to predict poverty levels in countries with limited survey and administrative data (Jean et al. 2016, Yeh et al. 2020). Although there are limitations to ML potential and applications (Kizilcec et al. 2020, McKenzie and Sansone 2019, Salganik et al. 2020), researchers have started to analyse how (and if) these algorithms can add value also when they are used to complement the judgement and skills of human experts (Stevenson and Doleac 2019). For example, once ML models identify high-risk individuals, caseworkers can then choose which types (or intensities) of programs ought to be targeted at which individuals – one way in which to address the contextual bandit problem (Athey 2019). To our knowledge, ML models have not yet been applied to predicting the risk of long-term income support receipt.

A second literature is the one that examines the factors explaining long-term income support receipt. Several relevant issues emerge when analysing this topic and they often stem from data limitations. First, welfare churn (i.e., repeated exit and re-entry into the welfare system) and welfare scarring (i.e., the current probability of income support receipt increases as a result of income support receipt in the past) are common phenomena: point-in-time measures or yearly snapshots of welfare receipt can fail to capture these dynamics (Bäckman and Bergmark 2011, Barrett 2000, Bhuller et al. 2017, Tseng and Wilkins 2003). Second, non-take-up of income support payments among eligible individuals is a significant issue and using data that excludes non-recipients fails to shed light on their income and employment outcomes (Bitler et al. 2003, Currie and Grogger 2001). And last, the influence of personal characteristics (such as age, sex, ethnicity, household structure) has been shown to interact with the impact of structural factors (such as geography, policy reform, macroeconomic conditions) in explaining entrenched welfare receipt (Bradbury and Zhu 2018). Therefore, accurately and flexibly modelling relationships between individual characteristics and long-term income support receipt requires larger data combined with a richer set of covariates.

This study bridges the innovations made in these two areas of research by developing predictive ML models of long-term income support receipt. We use novel data of the full population of Social Security System enrolees. The dataset is called "Data On Multiple INdividual Occurrences" (DOMINO). These data have several key benefits. First, they are high frequency with daily information of income support receipt status from 2000 to 2019. These repeat observations enable an analysis of the dynamics of welfare exit and entry. Second, they include both welfare recipients and individuals receiving other government transfers, so they also include individuals who did not receive income support payments. Third, they include rich information on over 32 million persons who had contact with the social security system anytime from 2000 to 2019, therefore representing a large dataset ideal for calibrating ML algorithms. Fourth, in contrast with previous ML applications relying on survey data and/or data that are not directly available to practitioners or relevant to their day-to-day tasks, caseworkers and front-line agency staff already have access to these administrative data. This allows us to validate our algorithms with data that are representative of the population of interest and makes it easier to integrate the ML algorithms developed here in future decision-making processes. There is also evidence showing that administrative data tend to outperform behavioural data and lessen privacy-invasion concerns (Bjerre-Nielsen et al. 2021).

We argue that our estimated ML models applied to these novel and rich data can be a large part of the solution to the resource allocation problem faced by welfare agencies. Other researchers have also emphasised the benefits of using ML in combination with large administrative data to identify at-risk individuals (Van Landeghem et al. 2021). Indeed, we first show that our estimated ML models significantly improve predictive accuracy compared to simpler predictive models or early warning systems currently in use. Specifically, our simulations show how ML algorithms can increase the $R^2$ of models predicting the proportion of time individuals will be on income support by at least 22% (or 14 percentage points) compared to alternative models. As a result, our machine learning algorithms can improve the detection of long-term income support recipients accruing a welfare cost nearly AUD 1 billion higher than individuals identified in the current system.

Second, we argue the predictive models can improve decision-making by systemising the process of identification, helping to promptly identify at-risk individuals using only short-term variables, potentially even as soon as they register with the system. This is especially useful when there has been no prior familiarity with the registrant's situation. In this way, our predictive models can reduce workload pressures on caseworkers (because they act as automatic screening devices) and, by avoiding an arbitrary selection of predictors or subgroups, have the potential to detect and thus prevent conscious and/or unconscious biases (Kleinberg et al. 2020, Kretsedemas 2005, McBeath et al. 2014, Pentaraki 2019), as well as "cream-skimming". Similarly, in contrast with heuristic evaluations and predictions relying solely on caseworker's expertise and training, these ML predictions can be provided to caseworkers without requiring them to undergo any extra training on ML. Furthermore, these improvements can be made at low-cost. In effect, when our predictive models are paired with caseworker expertise, they can enable a superior assignment policy.

After having identified at-risk individuals, this paper illustrates the application of unsupervised ML to cluster such individuals into different groups based on their observable characteristics. Clustering at-risk individuals has two advantages. First, it emphasizes that these individuals are not a homogeneous group: the ML algorithm may classify some individuals as at-risk because of their disabilities, or because they have caring responsibilities, while others may be predicted to be at risk because of their migration status or age. The latter group would likely require different intervention programs from the former. ML can therefore be used to identify individuals at-risk, and complement separate or subsequent causal analyses in designing treatments appropriate for each sub-population. Second, it can further supplement causal analyses by identifying groups based on complex interactions between background characteristics. This can improve upon existing analyses that focus on one dimension to form the sub-groups (e.g., age, sex, or race), and thus provide the basis for examining more intricate and multidimensional heterogeneous treatment effects in causal analysis such as through a Randomized Control Trial (RCT).

Relatedly, while our paper is focused on *prediction*, it can be seen as a first step to identify and characterise at-risk individuals: in a follow-up step, caseworkers can focus on a particular subgroup and use their expertise (potentially in combination with contextual bandit algorithms or findings from previous RCTs) to assign these individuals to specific programs with proven high treatment *causal* effects. We therefore believe that this paper provides an example of how ML can be used to obtain accurate predictions and complement – rather than substitute – standard econometric techniques aimed at estimating causal effects.

## 2. Institutional background

The primary purpose of Australia's social security system is to provide people with a "minimum adequate standard of living" (Australian Treasury 2010). The main class of benefits provided are called "income support payments" and they are targeted at individuals with no or low levels of income and/or assets. Generally, for these recipients, income support payments serve as their primary source of income. These welfare payments are provided on a regular basis and assists with basic living costs. The maximum annual income support amount in 2018 ranged from AUD 12,400 (for Unemployment Benefits) to AUD 20,700 (for Disability Benefits), whereas the median annual personal income was AUD 48,400 (ABS 2019b). Thus, those who are receiving an income support payment are a highly disadvantaged group.

There are six main categories of income support payments: (1) student payments; (2) unemployment payments; (3) parenting payments; (4) disability payment; (5) carer payment; and (6) age pension. These main income support payments are strictly means-tested. This means that a formal process is used to determine an individual's eligibility for payments. Entitlement is based on current (not previous) levels of income and assets. Cash transfer amounts reduce when earnings and assets increase. This targeted model applies to all the income support payments; however, the income threshold and taper rates (i.e., how much welfare transfers decrease when earning and assets increase) vary depending on the type of income support payment. For example, Unemployment Benefits have one of the highest taper rates at 50-60%, which means that benefits cut out as soon as recipients earn roughly 24,400 AUD. The taper rates for other benefits are lower at 40-50%. Unemployment Benefits in Australia are flat-rate payments provided to any individual who is currently unemployed,[4] conditional on them satisfying activity test requirements. It is important to note that Unemployment Benefits are distinct from Unemployment Insurance because eligibility for the former does not depend on previous earnings.

This strictness of the income and asset test gives the Australian welfare system the title as one of the most targeted systems in the OECD. For example, in 2005, the average share of transfers received by the poorest population quintile as a ratio of the share received by the richest quintile was 2.1 across the OECD but as high as 12.4 for Australia. In other words, "the poorest 20 per cent of the [Australian] population [received] more than 12 times as much in social security benefits as the richest quintile" (Whiteford 2010).

For the most part, the welfare system aims to support people in immediate need. It does this by providing highly targeted income support payments. Nearly a fourth of the Australian population receive these payments. However, the Australian government also provides financial support through the social security system that is not considered an income support payment. Other payments available through the social security system include: payments that are designed to assist families with the cost of raising children, such as family tax benefits, maternity leave benefits, supplementary payments for the main income support benefits; and rent assistance. As some of the non-income support payments are relatively untargeted, the social security system includes cross-sections of the population that are not necessarily financially disadvantaged. As an example of the composition of recipients in the social security system, in any one fortnight in 2018, about 5 million Australians received an income support payment while approximately 855,000 families received family tax benefits (Whiteford 2018). In total, social security payments equalled AUD 112.4 billion in the 2017-2018 fiscal year (AIHW 2019b).

While the Australian social security system provides for a minimum standard of living with payments targeted to people who do not have the means to support themselves, this goal is balanced with the aim to encourage self-reliance and reciprocity by including activity tests, monitoring and sanctions (Klapdor 2013). For example, to remain qualified for unemployment payments, recipients are required to actively look, plan and prepare for work in the future. Failure to comply with these mutual obligation requirements can result in loss of payments. Currently, the government contracts out all employment assistance programs for unemployed people to for-profit and not-for-profit providers under the "Job Services Australia" program. Providers are paid according to a combination of service inputs, investment in work experience and training, and importantly, funding is linked to employment outcomes. Researchers who have evaluated the Job Services Australia program have found evidence of "cream-skimming" by caseworkers, such as targeting employment activation programs at those for whom outcomes are easier to achieve - since payments for providers are tied to outcomes (O'Sullivan et al. 2019). ML models, by providing risk scores for each individual, can provide detailed guidance to the government about which families are most at-risk of long-term unemployment. This information can then inform payment structures for contracted-out employment services. In this case, ML can serve as a critical governance and auditing tool for government-employment service provider relations.

---

[4] New migrants are ineligible for Unemployment Benefits within four years of arriving in the country.

The government also has additional education and training programs and mentoring services specifically aimed at reducing long-term receipt. For example, in Australia in 2016, as part of a 92-million-dollar initiative called "Try, Test and Learn", the government targeted early intervention programs at three high-risk groups of individuals: young carers age 24 or under, young parents age 18 or under, and students receiving study-related income support payments. These groups were selected based on an actuarial valuation of their estimated lifetime welfare costs (Department of Social Services, 2015, Price Waterhouse Coopers, 2016; 2019). Other considerations included the adequacy of available support services for these groups and their expected responsiveness to new models of policy intervention. The initiative funded one-on-one tailored support, including short-term education, workshops, mentoring, work placements, and self-employment/business start-up initiatives. In this context, ML models can help to further refine the process of identification of high-risk individuals, especially those within the three identified broad groups who may have different risk profiles.

## 3. Data

### 3.1 Population

Australian federal social security records from 2000 to 2019 form the basis of our dataset. All social security payments are administered by the national welfare agency called Centrelink. There are over 32 million persons in these data who had any contact with the Centrelink system between 2000 and 2019. All registrants are 15 years or above as this is the minimum age of eligibility. The financial circumstances of these registrants vary greatly: some have high levels of financial needs, such as those who are in receipt of highly-targeted income support payments; others have higher incomes and register with the social security system because they receive one of the non-income support payments described in the previous section, such as one-off government bonuses or cost-of-children payments.

Each individual is tracked over time on a highly frequent basis. For our main variable of welfare receipt status, we know the precise start and end date associated with payment receipt. A key advantage of this data structure for our study is that we can construct a precise picture of the duration and dynamics of welfare receipt over a long period of time.

The reason we observe such high-(daily)-frequency data is because income support (or welfare) payments are highly targeted. This means that recipients' eligibility for payments are assessed regularly and recipients are required to report changes (such as to relationship status, earnings or living conditions) within 14 days of the change. The start and end dates of payment receipt are then recorded in our data. Specifically, recipients are required to report their financial circumstances and living arrangements to Centrelink on a regular (bi-weekly) basis by filling in a 14-34-page form that elicits information about the recipients' (and if applicable, partners') basic information (name, address, contact details, gender, date of birth, ethnicity, language, citizenship, arrival information), marital status and relationship event history, demographic information about their dependent children, accommodation details, employment and study details.

### 3.2 Sample

Our first step is to identify the universe of individuals who received any type of payment from Centrelink in 2014, and who were aged 15 to 66 on 1 January 2014. On average, each person in this sample incurred approximately AUD 160,950 in welfare expenditures over the 2000-2018 period.

We have chosen to focus on the working-age population since this is the target group for programs aimed at preventing long-term welfare dependency. Our next step is, for computational reasons driven by the power of the server where the data are stored, to draw a random 1% sample (50, 615 individuals) from this population.[5]

Furthermore, we have chosen to focus on individuals in the social security system in 2014 (base year). The resulting sample is not representative of the whole Australian population (Table 1). It disproportionately captures individuals with low- to middle-incomes. The highly targeted nature of Australian welfare programs means low-income individuals are over-represented. Furthermore, the non-income support payments, such as the cost-of-children payments, are provided to primary carers of children. This means there is a higher percentage of females in our sample since they continue to be over-represented among primary carers of children (as well as among individuals eligible for other carer-related payments). Nevertheless, it is worth emphasizing that this sample is appropriate for our goal of predicting welfare dependency given its focus on vulnerable individuals.

We then calculate an individual's welfare receipt intensity from 2015 to 2018, inclusive. Note that we ignore the data from 2000 to 2013. This allows us to estimate the machine learning algorithms in a period after the Global Financial Crisis effects have washed out. We have also chosen to exclude the year of 2019 because the end of our observation window is 14 October 2019. Partial inclusion of 2019 data points could bias our results if there are strong seasonal patterns associated with welfare receipt.

**3.3 Dependent (outcome) variable**

We calculate the intensity of welfare receipt as the proportion of time an individual received an income support payment during the four-year period spanning 2015 to 2018: i.e., the number of days they received an income support payment from 1 January 2015 to 31 December 2018 divided by the total number of days over this period. We have chosen this four-year period over which to calculate the duration of welfare receipt because it is long enough to identify long-term welfare dependence but also short enough so that we don't obscure the individual's needs at different stages of the life cycle.

---

[5] Note that the Australian population aged above 15 in 2014 was approximately 19 million (ABS 2014). Approximately seven million individuals were registered in the Centrelink system in 2014 (our base year). Of the 2014 Centrelink registrants, roughly five million were aged 15-66 years. A 1% sample gives us our final sample of 50,615 individuals.

Our measure of welfare receipt intensity is consistent with how welfare reliance has been conceptualised by previous studies and in government reviews such as the most recent McClure review (Reference Group on Welfare Reform 2015). It also captures a more severe form of economic disadvantage than measures that only consider the receipt of income support at one point in time. This is important since receipt of welfare payments is not the same thing as being dependent on welfare – a point that has been well emphasised in the literature (Penman 2006).

Long-term welfare receipt is a prevalent issue in Australia. For example, in 2018, almost 3 in 4 income support recipients aged 18-64 had been on a payment for two years or more and among those receiving income support payments in 2009, more than half (56%) were receiving payments 9 years later in 2018 (AIHW 2019a). This issue is also increasing in importance. In 2018, 24.5% of unemployed people aged 15 and over had been looking for work for more than a year (annual average), increasing from 14.8% in 2009 (ABS 2019a).

Figure 1 plots the distribution of income support intensity. In Panel A of Figure 1, focusing on everyone in our sample, we see spikes in the proportion of people who never received income support (32.3% of the sample), and those who received income support consistently over the period 2015 to 2018 (36.7% of the sample).[6] In Panel B of Figure 1, we display the distribution of income support intensity once we exclude the two extreme cases. Here, we see a more uniform distribution across the spectrum of intensities with some concentration at the higher end.

As an extension, we also predict the outcome of "long-term unemployment benefit receipt". This proxies for long-term unemployment, which is particularly relevant given the recent COVID-19-induced recession and the attendant high rate of unemployment. Focusing on unemployment benefits is also useful because long-term receipt of this payment indicates a failure to achieve its primary purpose: to act as a temporary payment while people transition from unemployment to employment. Furthermore, as previously mentioned, our ML models can potentially fulfil a useful governance and auditing role by ensuring that individuals who are most at risk of being long-term unemployed are not unduly ignored through "cream-skimming" when third-party employment service providers make resource allocation decisions.

---

[6] This type of bi-model pattern is common across the entire period between 2000 to 2019.

Figure 2 plots the distribution of unemployment benefit intensity. Similar to the pattern of "any" type of income support receipt, we see spikes in the proportion of people who never received unemployment benefits (50.5% of the sample), and those who received unemployment benefits consistently over the period 2015 to 2018 (30.3% of the sample). This is illustrated in Panel A of Figure 2. Once we exclude the two extreme cases (Panel B of Figure 2), we see a higher concentration of intensities at the lower end.

### 3.4 Independent (input) variables

The DOMINO data capture a wide range of information on individuals. Based on data observed in 2014 (the base year), we include information on: demographics (sex, age, country of birth, and Indigenous status); household structure (parent status, number of children, and ages of children), government benefit receipt history (by benefit type); personal relationships (partnership status and marital status, as well as the duration and instability of relationship status); employment and underemployment (zero-hour contracts); work instability; location and residential mobility; housing; education; income and wealth.

A key benefit of the high frequency DOMINO data is the ability to model the dynamics of economic behaviour. For example, we include variables that capture the instability, variability, and intensity (number or duration) over time of income support history and employment. Examples include: exits and entries onto the income support system, annual fluctuations in the amount and duration of benefits, wages and earnings, hours of employment, number of jobs (including simultaneously held jobs), and changes in the working arrangements (day-of-week and hours per week) both within the same employer and across employers.

One limitation of the DOMINO data is a dearth of variables directly measuring the individual's non-cognitive ability or behavioural tendencies, such as risk preferences or forward-looking behaviour. We have included a range of proxy variables for these unobservable factors. For example, we include an array of (over 60) measures of intensity and volatility of past employment spells and income support history (see Online Appendix A). Such labour market history variables have been shown to act as good proxies for behavioural traits in predicting unemployment and reemployment success, absorbing the variation from traits such as years of education, self-reported job search, and conscientiousness (Van Landeghem et al. 2021). In addition, we include measures such as whether the individual was ever sanctioned or had their benefit cancelled because of non-compliance reasons. Indeed, one interpretation of being sanctioned may be disorganisation or delays (Banerjee and Duflo 2014) on the part of the registrant (to fulfil mutual obligations or activity test requirements, for example). However, we recognise that these measures are likely to reflect other factors and constraints as well (Klapdor 2013).

## 4. Methods

### 4.1 Simple and heuristic models

We begin by providing a benchmark for the ML performance and to proxy current methods used by practitioners when predicting individuals at-risk of welfare dependency. To do this, we estimate simple OLS models that predict the proportion of time an individual spends on income support between 2015 and 2018, as reported in Table 2. The most basic benchmark is a regression containing only a constant term (Model 1). Building on this, we evaluate the predictive power of demographic characteristics: sex (Model 2), education (Model 3), and age (Model 4). We then consider the predictive power of income support history – i.e., a series of binary variables indicating whether an individual had ever received an income support payment at any time in 2014, separately for different types of income support payments - in Model 5.

Another natural benchmark is to estimate a heuristic model, which includes variables that the past literature has identified as key drivers of welfare dependency such as sex, age, education level, parent-status, migration status, ethnicity, marital status, state of residence, and unemployment status (Model 6). We further enhance this model by adding the income support history from Model 5 (Model 7). In addition, given the emphasis in the past literature on local neighbourhood effects (Chetty et al. 2018), we include detailed geographical information in Model 8.

Finally, we test the potential predictive gain of ML models over the profiling indicators currently used by the Australian government to prevent welfare dependency (Model 9). As outlined in Section 2, these are indicators for the three groups with the largest estimated lifetime welfare costs:

young carers age 24 or under, students receiving a particular income support payment, and young parents age 18 or under.

**4.2 Machine Learning approach**

The heuristic models rely on a small set of variables and on methods like simple aggregation for data reduction. Yet, in practice, simple models are unlikely to represent the complex processes underlying welfare dependency. Fortunately, our baseline data enable greater flexibility because it contains a rich set of possible predictors. We include a wide array of these predictors and different functional forms for each predictor. For example, we include a full set of indicator variables for a recipient's country of birth, their marital status, employment and education status, along with other non-linear expressions (and complex interactions) between these variables (as described in detail in the Online Appendix).

Following Mullainathan and Spiess (2017), we do not drop redundant or aggregated variables since they could be useful to obtain better predictions with less complexity. For example, we include categorical variables recording the number of children in the household while also including an extensive set of binary indicators indicating family size and fertility progression. The net result of this is that we have approximately 1,800 possible predictors. Moreover, if we start considering interaction terms between some of these predictors, the number of variables can reach or even exceed the number of individuals in the data. In a standard OLS regression, this would not be possible.

By contrast, ML efficiently handles high-dimensional data. This allows us to identify previously undetected relationships between variables. A common issue with adding complexity to a model (in the form of more variables, more interactions, or more flexible functional forms), however, is overfitting. This issue is evident when an estimated model tightly fits a training sample but poorly fits new samples. The problem of overfitting cannot be easily solved by estimating simple models (e.g., with fewer variables subjectively selected) because this may result in models that underfit the data and exclude powerful predictors. For instance, if the true relation between y and x is quadratic, a linear model would be an under-fit (high bias), while estimating a $4^{th}$ degree polynomial would lead to over-fitting (high variance). ML addresses the bias-variance trade-off by carefully reducing the number of variables, as well as through a process called regularization. The latter keeps all the variables but reduces the magnitude of each coefficient. This works well when there are numerous variables that contribute to predicting the outcome in a statistically significant, albeit economically modest, fashion. We have used three different ML algorithms which employ one or both of these methods for dimension reduction.

Estimating a range of ML models (as we do) is advisable because each have their benefits and drawbacks. As noted by Athey and Imbens (2019), there are no formal results that shows that one ML approach is better than another. Therefore, the choice of which algorithms to use can often be rather arbitrary. We choose to estimate different classes of ML models, including LASSO, Support Vector Regression (SVR) and Boosting (reported in Table 2 Models 10-12). This allows us to evaluate algorithms which have different levels of flexibility and interpretability. LASSO is straightforward to explain given its similarity with OLS and can be easily interpreted. SVR offers extremely flexible functional forms. Similarly, Boosting is a very flexible tree-based algorithm which can account for a large number of possible interactions among the inputs. Indeed, in this specific application we allow up to 6-way interactions between input variables in our Boosting algorithm. We have briefly summarised each of these algorithms in the Online Appendix. A more comprehensive review of the tools available to practitioners is provided by, among others, Hastie et al. (2009), Mullainathan and Spiess (2017), Athey and Imbens (2019).

We follow the recommendation of Mullainathan and Spiess (2017) and split the data into two sub-samples. A training sample (80% of the data) is used to calibrate and estimate the algorithm under each of the ML methods. Out-of-sample performance is reported using the hold-out sample (the remaining 20% of the data). A detailed step-by-step description of our calibration procedure is available in the Online Appendix.

In many cases, a single algorithm does not perform as well as a combination of methods. Different machine learning algorithms may capture different features of the data. Therefore, combining these algorithms – as pioneered by Bates and Granger (1969) - may potentially lead to a superior performance (Athey and Imbens, 2019). We have aggregated in Table 2 Model 13 the predictions from the three main ML algorithms (LASSO from Model 10, SVR from Model 11, and Boosting from Model 12) using weights obtained through running a linear regression of the outcome on these predicted values, as described in Mullainathan and Spiess (2017).

## 5. Results

Table 2 shows the out-of-sample performance of the aforementioned models when predicting the proportion of time an individual receives income support between 2015 and 2018 using their available information from 2014. For each model, we report the Mean Squared Error (MSE) to compare the predictive accuracy of different models. The MSE is a widely accepted criterion to measure the performance of models that predict continuous variables. Given its intuitive appeal, we also report the square of the Pearson correlation coefficient, i.e., the correlation between the actual and fitted dependent variable. This is equivalent to the $R^2$ in linear least squares regressions. All these goodness-of-fit measures are estimated out-of-sample using a holdout sample. We use a separate training sample to estimate and calibrate the models. In-sample performances are reported in Table B4 in the Online Appendix.

In Models 2-4, we report that demographic characteristics have limited predictive power: OLS models including only information on sex or education have an out-of-sample $R^2$ of 3% or lower (Models 2-3). The OLS model using age as the predictor explains a higher percentage of the variability in the outcome variable (17.4%, Model 4). This is to be expected given the age requirements in several income support programs. Despite the improvement in predictive performance over Models 2-3, the $R^2$ remains low in absolute terms.

In Model 5, we show that income support history is strongly correlated with future welfare dependency. Specifically, a simple OLS model that includes binary indicators for any income support payment (separately for different types of payments) received in 2014 can explain almost 60% of the variability in the proportion of time on income support in the four subsequent years.

In Model 6, we find that the heuristic model does not improve performance in a substantive way. This is surprising as the set of predictors of demographic characteristics, location, employment and education status is commonly used by economists to explain economic behaviour. Further adding this set of predictors to Model 5 (i.e., with the income support history) only leads to a small reduction in the MSE. This suggests that our indicators of income support history are stronger predictors of future welfare receipt intensity. We obtain similar conclusions when we expand the set of controls to include detailed geographic information (Model 8). In addition, we show that the three at-risk groups identified in the "Try, Test and Learn" program have low predictive power when modelled in an OLS specification (Model 9).

Importantly, the results in Table 2 indicate that using ML to exploit a larger set of information already available in administrative data (that are accessible by caseworkers) lead to performance gains (Models 10-13). Consequently, ML predictions, once integrated into the current identification approaches used in the "Try, Test and Learn" program, could act as a first step in targeting early interventions at the most at-risk individuals. These improvements are consistently achieved irrespective of the particular ML algorithm implemented. The stability of the performance across algorithms is both reassuring and surprising given the contrast in their data-fitting mechanics (in terms of the flexibility allowed for functional forms and interaction terms).Combining all three algorithms (LASSO from Model 10, SVR from Model 11, and Boosting from Model 12) in Model 13 further improves prediction performance.[7] It is important to note that we obtain this out-of-sample predictive accuracy without overfitting the data (as evident in Table B4) and by using a 1% random sample of the full set of observations. In our context, it suggests that further sample size augmentations (and the attendant computational costs) are unnecessary to properly calibrate these ML algorithms.

The Ensemble method, compared to the benchmark models (Models 1-9), performs significantly better. Specifically, its out-of-sample MSE is less than one-fourth of the corresponding MSE of the basic benchmark model using only a constant (Model 1). Furthermore, its MSE represents a 42% reduction from the MSE of Model 5 with only the income support history. Bootstrapped confidence intervals are also tight around the mean MSE, suggesting that all comparisons in predictive performance between OLS and ML models are significantly different from zero. Similarly, the out-of-sample $R^2$ jumps to more than 76%, almost a 14-percentage point (or 22%) increase – again, at no extra cost – compared to the OLS model with the highest out-of-sample $R^2$ (Model 8). According to our back-of-the-envelope calculations of the total annual accrued welfare costs, the individuals identified in the ML model accrued an additional welfare cost of AUD 0.99 billion compared with the comparably sized (three) groups identified in the actuarial model. To give a sense of the magnitude, this represents roughly 10% of the total annual unemployment benefit expenditure (AIHW 2019b).

---

[7] This improvement is achieved despite the correlations of the out-of-sample predictions between the LASSO, SVR, and Boosting models being very high (between 0.97 and 0.99), suggesting that even under these conditions there are additional benefits of weighting and combining different algorithms.

Given the results in Model 5 Table 2, agencies lacking the time or resources to develop and apply more sophisticated models could identify - with a medium level of accuracy - individuals already in the system who are at risk of relying on income support for an extended period of time using only past income support payment information. Alternatively, one could focus on the top predictors from LASSO (Model 10 Table 2) and Boosting (Model 12 Table 2): Table 3 lists the most powerful predictors selected by these two algorithms. While most of these variables are not surprising, a subset of them – such as income fluctuations, number of residential moves, and failure to complete means tests - is not routinely used in simple early warning systems. As shown in Models 1-3 in Table 4,[8] using these selected inputs in an OLS model would yield high out-of-sample performance. Therefore, while ML algorithms reach the lowest out-of-sample MSE, insights from these algorithms can be easily incorporated into simpler modelling frameworks such as an OLS model. This is especially useful if technical, political, or administrative barriers prevent ML algorithms from being implemented at a larger scale.

## 6. Extensions and robustness checks

Table 4 reports additional extensions and robustness checks. Despite documented large economic inequalities between native and non-native populations (Markham and Biddle 2018), Indigenous status has limited predicted power in this context (Model 4). Similar conclusions are reached when focusing on country of birth (Model 5). In Model 6, to investigate how much income support dependency is affected by local factors, we further extend our set of detailed geographical inputs to include almost 2,500 postal codes, but we do not improve substantially upon the model that includes only information on income support history (Model 5 in Table 2). Including indicators of parental welfare receipt - a factor that has been shown to increase the likelihood of one's own welfare receipt (Cobb-Clark et al. 2017, Dahl et al. 2014, Dahl and Gielen 2020) - to the heuristic models has only a marginal effect on out-of-sample performance (Models 7-8).

Finally, Models 9 to 11 use the same variables used in the heuristic Model 6 in Table 2, but as inputs in a Fractional Outcome Regression (FOR) – Probit specification (to account for the clumping in the outcome variable at 0 and 1, as shown in Figure 1), and as inputs in a Boosting algorithm and a LASSO algorithm rather than an OLS model. The out-of-sample $R^2$ increases from 30.6% to 31.2% for the Fractional Outcome Regression; and it increases to 37.3% for the Boosting algorithm. This suggests that interaction terms and non-linearities do play a role in this context, but they are not enough to reach the same performance as the OLS model with the income support history variables or the ML algorithms with the full set of inputs. As shown in Model 11 Table 4, LASSO with the same inputs as the heuristic model (Model 6 in Table 2) does not improve upon the OLS model.[9]

---

[8] We report in-sample performance in Table B5. It is worth emphasizing that, as discussed in Section B of the Online Appendix, Boosting weights and combine different predictions from several (tree-based) classifiers, and at each iteration it uses only a random subset of the training data to build such trees. Therefore, this algorithm is more robust to outliers than LASSO, and the resulting list of top predictors is less likely to be influenced by specific observations or values in the training sample.

As an additional sensitivity, in Model 12 in Table 4, we present results for a LASSO model where we manually add interaction terms to the main LASSO model (from Model 10 in Table 2).[10] The added gain of including additional interaction terms (two-interactions between the top 20 predictors from the LASSO – Model 10) is minor.[11] For example, the out-of-sample $R^2$ in Model 12, Table 4 is only 1.3 percentage points higher than the LASSO in Model 10, Table 2 where we do not include any interaction terms. The predictive gains of Model 12, Table 4 over and above the SVR (Model 11, Table 2) and Boosting (Model 12, Table 2) are even smaller, pointing to the fact that these latter two algorithms automatically allow for highly flexible functional forms.

We exclude those who were on income support 100 percent of the time from 2011 to 2014 in Models 13-15. Comparing Model 15 with that of Models 13 (Heuristic model as in Model 6 Table 2) and 14 (same predictors as the heuristic model plus income support history as in Model 7 Table 2), we see that the LASSO algorithm substantially improves on the predictive performance of the OLS models.

Individuals with a historical pattern of high reliance on income support may be more difficult to mobilise into employment. This can be because of the potential scarring effects of long welfare receipt durations and/or because the very factors that lead these individuals to rely on welfare for long durations also prohibit them from finding employment. For example, approximately 61 percent of individuals who were on income support over the entire period of time from 2011 to 2014 were either receiving Disability Support Pensions themselves or caring for a family member with a disability (and thus receiving Carers Payment). In this set of sensitivity analysis, our motivation for excluding the individuals who are long-term welfare dependent from our sample is because conventional early intervention programs of a labour market activation nature may be less effective when targeted at these individuals. In fact, recipients of Disability Support Pensions and Carers Payment are not required to fulfil the same participation requirements as those receiving Unemployment Benefits. Reflecting this, the former group are not even referred onto Job Service Australia case managers who administer labour market activation interventions only for those who are in receipt of Unemployment Benefits.

---

[9] This can be explained by the fact that LASSO only penalized one variable in this simulation: i.e., it chose 34 of the 35 variables in the heuristic model.
[10] We restricted the two-way interactions to the top 20 predictors for computational reasons.
[11] Table B6 lists the top 10 predictors from the LASSO with interactions model (Model 12, Table 4). Note there is a high correlation in the base predictors chosen in this model with those chosen in Model 10, Table 2, suggesting stability in the importance of these key predictors.

It is important to highlight that we nominate to include the high-reliance group in our main analysis sample because policymakers explicitly aim to assist both those who are currently long-term welfare dependent, as well as those who are at-risk of it. Identifying both these groups in our algorithms does not preclude a differentiated approach in the type of assistance or intervention program that is targeted at them. For example, identifying those who are most likely to be long-term welfare dependent because of enduring health issues warrants the more generous program of monetary assistance and lower participation requirements that are currently in place; whereas those who are at-risk of being long-term jobseekers may be better helped with labour market activation strategies. Furthermore, within the group of long-term jobseekers, some of them may require more intense case management (or different forms of activation strategies) than others. If individuals in the high-reliance groups receive the generic support programs (or worse, are systematically ignored by case managers because of the aforementioned "cream-skimming" issues), then the chances of mobilising them into employment may be low. In summary, individuals who are identified as being at-risk of long-term welfare receipt are a heterogenous group and are likely to have a different set of needs. The treatment strategies aimed at them should be tailored accordingly. We explore this in more detail when we estimate unsupervised ML models in the next section.

Finally, in Table 5, we have estimated another set of results using the outcome of long-term unemployment benefit receipt rather than the more general long-term welfare receipt analysed so far. The ML models perform slightly better than the benchmark models. For example, the out-of-sample $R^2$ increases from 83.1% to 86.2% from Model 8 to Model 9, and the bootstrapped 95-percent confidence intervals for the MSE do not overlap. It is important to note that while the gain from ML seems to be larger when the algorithms are applied to the whole sample of welfare recipients in Table 2 (and even if the predictive gain from LASSO in Table 5 is only 3 percentage points) the cost of calibrating an ML model on existing datasets is negligible. This makes it worthwhile to estimate ML models over simpler OLS models also in this context.

## 7. Unsupervised machine learning

As clear from our discussion in the previous sections, individuals who receive income support intensively are unlikely to be a homogenous group. There may be more narrowly defined populations because of two factors. First, the Australian welfare system is structured to provide different payments to address distinct needs in the population (for example, disability benefits are provided to those who have severe health issues; unemployment benefits are provided to those who are unemployed; and parenting benefits are provided to primary carers of young children). Reflecting this, different payments have varying eligibility criteria, as well as different levels of generosity. This means individuals receiving different income support payments are likely to be receiving payments for varying lengths of time simply because of their pre-existing characteristics and conditions. Second, amongst those receiving the same payment, the reason/s for which an individual is receiving a payment can differ. For example, unemployment payment recipients comprise of individuals who are transitioning from study to employment and those who are discouraged job-seekers.

Thus, if case workers and welfare administrators want programs to address individual needs and circumstances, they may want to design different treatments. Unsupervised ML can be a useful first step in undertaking this task because it identifies the distinct clusters within the population of long-term welfare recipients. In addition, it can partition clusters according to complex interactions between different variables. This is important because individuals may be at risk of long-term welfare receipt due to a complex set of circumstances, not just due to one dimension alone, such as based on their age, ethnicity, or gender. In other words, this section acknowledges that long-term dependency on income support is a multidimensional issue: different factors may underlie prolonged receipt patterns. This is similar to the multidimensional approach advocated in poverty studies (Alkire and Foster 2011). This section thus shows how individuals predicted to be at risk can be divided into different subgroups using unsupervised ML.

So far, we have focused on predictive models, i.e., on supervised ML. Indeed, supervised algorithms are provided with a certain number of "right" answers, i.e., actual *y* associated with a certain *Xs*, and are asked to produce other correct answers, i.e., to predict new *y* given other combinations of *Xs*. On the other hand, unsupervised learning algorithms derive a structure for the data without necessarily knowing the effect of *x* on *y*.

The starting point is the prediction obtained using the LASSO algorithm in Table 2 (Model 10). We have then used the top variables selected by LASSO (as reported in Table 3) combined with the indicators from the heuristic model and income support history variables in Table 2 (Model 7) to divide the individuals predicted to be at risk into different groups by means of a hierarchical clustering algorithm (Stata 2019). More formally, cluster analysis specifies whether the joint density of the observable variables *X* can be represented by a mixture of simpler densities representing distinct groups of observations. Conceptually, the hierarchical clustering algorithm can be summarized as follows: initially there are *n* distinct groups, one for each observation; in the next step, the two closest observations are merged into one group, thus resulting in *n-1* groups; after that, the closest two groups are merged together, producing *n-2* groups. This process continues until all the observations are merged into one large group. Therefore, the output of this algorithm is a hierarchy of groupings from one group to *n* groups. As explained in Sansone (2019), the Caliński and Harabasz pseudo-F index and the Duda-Hart Je(2)/Je(1) index with associated pseudo-$T^2$ can help analysts to select the best number of groups.

We find five clusters most tightly partitions the feature space. Table 6 shows the summary statistics for these predicted at-risk individuals.[12] For comparison purposes, we have also reported the summary statistics for the individuals who are predicted to never receive income support (last column). Based on these summary stats, we see some clear similarities between the four groups of long-term welfare recipients. For example, all of these individuals had a higher incidence of tenuous employment as defined by working in more than one job at one point in time, being more likely to be Indigenous, and comprising of individuals aged at the extreme ends of the distribution (for example, below 30 and/or above 55). Also, many of them were less likely to have been married in 2014 (our base year).

---

[12] For simplicity, only the key variables have been reported Table 6. Summary statistics for the whole set of predictors are reported in the Online Appendix (Table B8). We only report summary statistics for four of the five clusters since one of the clusters includes only 5 observations. For this small group, reporting their summary statistics would compromise data confidentiality rules and would fail to provide any meaningful conclusions about the groups' characteristics.

Clear differences in the background characteristics also emerge between the groups. For example, people who were more likely to receive disability benefits themselves or receive a payment on behalf of a disabled family member are overrepresented in Group 1, while Group 2 has a higher percentage of immigrants. Group 3 is overrepresented by Indigenous individuals, younger people, and those who receive unemployment benefits. Finally, Group 4 is more likely to be composed of female recipients with children.

This result emphasizes the importance of not pooling together all individuals identified as at risk. In fact, placing in the same program individuals with disabilities side by side with young Indigenous individuals or single mothers may not be the most efficient approach and may actually result in negative program effects.

## 8. Conclusions

In this paper, we show that ML algorithms applied to data already available to caseworkers can significantly improve predictive accuracy compared to OLS models that use a smaller set of variables or compared to currently used early warning systems. The ML predictions can be applied to potentially reduce workloads for caseworkers and to supplement their expertise in the process of identifying high-risk individuals. We find evidence that using the better-performing ML models, as opposed to the currently used early warning indicators, could lead to substantial savings in public spending because individuals identified by the former accrue on average higher welfare expenditures than those identified by the latter. We then show how unsupervised ML can be exploited to identify sub-populations among high-risk individuals.

Several caveats are worth noting. First, although we have used some of the most popular ML algorithms, it is possible that more advanced algorithms or extensive grid-searches could further improve performance. In addition, these algorithms have been designed to be applied to large data sets with millions of observations and thousands of potential inputs. Even if administrative data have extremely large numbers of observations, the number of potential inputs is often limited. For instance, we did not have any information on where the individual went to college, the individual's risk preferences, their non-cognitive skills, or the full record of their parents' education and employment history. We should also emphasise that our simulations are based on data collected during a period of stable or improving macroeconomic conditions; ML algorithms would need to be retrained using data from economic downturns in order to improve external validity claims, especially when predicting welfare dependency among individuals affected by the Covid-19 pandemic. Despite these limitations, the findings in this paper suggest that researchers and practitioners could obtain more precise predictions – at no extra costs - by exploiting the available datasets.

Second, the predictions in this paper are different from treatment effect estimates that a policy-maker may use to optimally allocate resources (Athey 2017): such agents would be more interested in knowing which individuals would benefit the most from a welfare transfer. For example, in the "Try, Test and Learn" program in Australia, one consideration in selecting the three at-risk groups was how responsive they would be, in a causal sense, to new interventions. But even in such cases, accurately identifying a pool of high-risk individuals may be a useful first step. Furthermore, by systemising the screening process, it can potentially limit the impact of explicit and implicit biases or "cream-skimming", as well as allow caseworkers to focus on the most critical welfare recipients. In other words, supervised ML can be used in the first stage to identify high-risk individuals, while unsupervised ML can divide these individuals into subgroups. These findings can then complement and support separate or subsequent causal analyses to inform policy-makers about the appropriate intervention/s for high-risk groups. In addition, from a social welfare perspective, policymakers may prioritise allocating resources to the most vulnerable individuals even if it does not necessarily aim to maximise the marginal gains from a treatment program.

Future research could explore how to use the predictions from these algorithms as a preliminary step in randomised control trials (Duflo 2018). For instance, these algorithms could help researchers to identify their population of interest, e.g. the subset of individuals most at-risk, and then design an RCT to find the appropriate treatment to support these welfare recipients. Chandler et al. (2011) provide an early example of incorporating predictive models in public policy programs targeting specific sub-populations. In this context, ML techniques known as multi-armed or contextual bandits can regularly update algorithms with newly harvested information (on registrants) and recommend personalised treatments (Athey, 2019). Finally, future research could elucidate how ML algorithms ought to be effectively combined with human expertise, including an understanding of the contexts in which one should rely heavily on ML algorithms and those in which we ought to more heavily rely on humans (Raghu et al. 2019). Advances in this knowledge base can ultimately improve future resource allocation decisions.

**Figure 1: Density of proportion of time on income support between 2015 and 2018.**

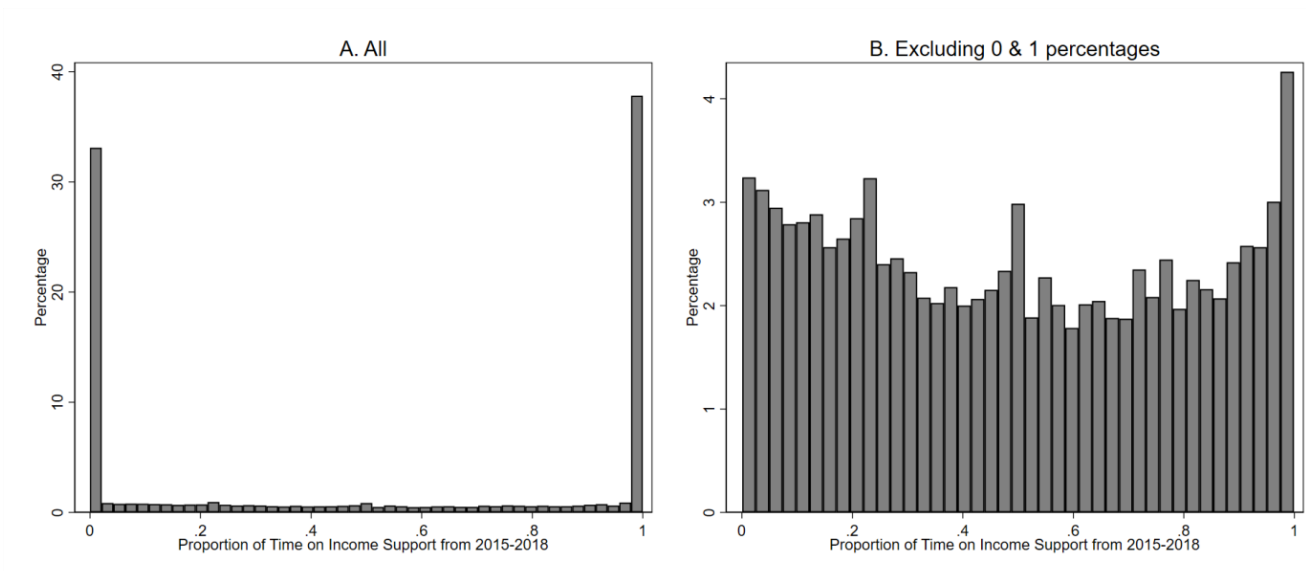

*Notes*: These plots show the distribution for the proportion of time individuals received an income support payment between 2015 and 2018. Individuals with no income support or always on income support are excluded in Panel B.

**Figure 2: Density of proportion of time on unemployment benefits between 2015 and 2018.**

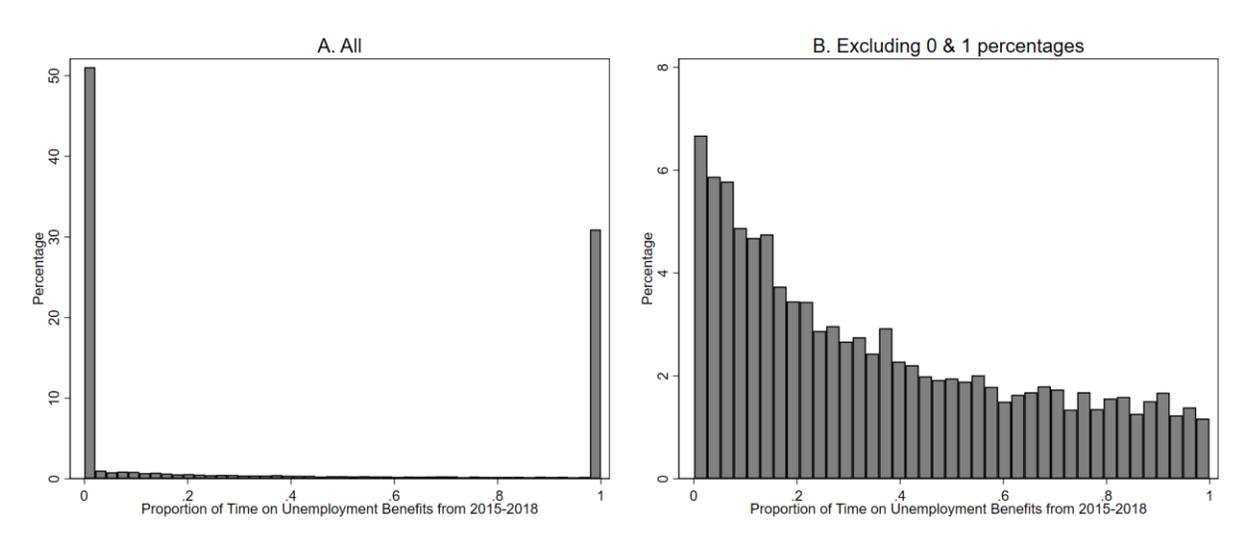

*Notes*: These plots show the distribution for the proportion of time individuals received an unemployment benefit between 2015 and 2018. Individuals with no unemployment benefit or always on unemployment benefits are excluded in Panel B.

**Table 1: Summary statistics comparison between Census and DOMINO data.**

|  | Census | DOMINO |
|---|---|---|
| **Demographics** | | |
| Age (mean) | 39.825 | 40.954 |
| Female | 0.505 | 0.620 |
| Ever a parent (with at least one child) | 0.291 | 0.468 |
| Indigenous | 0.027 | 0.057 |
| Australian-born | 0.680 | 0.721 |
| Immigrant | 0.320 | 0.279 |
| **Education** | | |
| Year 12 and below | 0.415 | 0.549 |
| Certificate I/II and below | 0.416 | 0.785 |
| Certificate III/IV and below | 0.619 | 0.839 |
| Diplomas and below | 0.725 | 0.904 |
| Bachelor and below | 0.915 | 0.986 |
| **Marital status** | | |
| Single | 0.430 | 0.542 |
| Married | 0.431 | 0.361 |
| De Facto | 0.133 | 0.097 |
| Separated | 0.025 | 0.182 |
| Divorced | 0.058 | 0.038 |
| Widowed | 0.011 | 0.018 |
| **State of Residence** | | |
| New South Wales | 0.318 | 0.318 |
| Victoria | 0.255 | 0.248 |
| Queensland | 0.200 | 0.211 |
| Western Australia | 0.070 | 0.087 |
| South Australia | 0.107 | 0.080 |
| Australian Capital Territory | 0.021 | 0.012 |
| Northern Territory | 0.011 | 0.011 |
| Tasmania | 0.018 | 0.027 |
| **Weekly Personal income (Annual income)** | | |
| Nil (or negative) income | 0.106 | 0.134 |
| $1-$149 ($1-$7,799) | 0.046 | 0.467 |
| $150-$299 ($7,800-$15,599) | 0.067 | 0.134 |
| $300-$399 ($15,600-$20,799) | 0.059 | 0.052 |
| $400-$499 ($20,800-$25,999) | 0.060 | 0.021 |
| $500-$649 ($26,000-$33,799) | 0.069 | 0.012 |
| $650-$799 ($33,800-$41,599) | 0.077 | 0.010 |
| $800-$999 ($41,600-$51,999) | 0.089 | 0.005 |
| $1,000-$1,249 ($52,000-$64,999) | 0.093 | 0.003 |
| $1,250-$1,499 ($65,000-$77,999) | 0.065 | 0.001 |
| $1500 or more ($78,000 or more) | 0.187 | 0.000 |
| Not stated | 0.082 | 0.295 |

*Notes*: Samples for both Census and DOMINO are restricted to individuals aged between 15 and 66. The difference in means (Columns 2 and 3) are always statistically different from zero at a 5% significance level, according to two-sample T-tests. Exceptions only include the binary indicators for New South Wales and the Northern Territory. Census data taken at 9th August 2016. Statistics for the education levels are calculated as a fraction of non-missing observations. Missing rates for education in Census and DOMINO are 12% and 53%, respectively. Statistics for marital status are calculated as a fraction of non-missing observations. Missing rates for marital status in Census and DOMINO are 11% and 0.1%, respectively. All variables are described in Online Appendix A.

**Table 2: Prediction (out-of-sample) accuracy for time on income support.**

|  |  | Model | Predictors | MSE Mean | MSE C.I. | $R^2$ |
|---|---|---|---|---|---|---|
| Simple models | 1 | OLS | Constant | 0.203 | [0.202; 0.205] | 0.0% |
|  | 2 | OLS | Sex | 0.202 | [0.200; 0.203] | 0.9% |
|  | 3 | OLS | Education | 0.197 | [0.195; 0.199] | 3.0% |
|  | 4 | OLS | Age | 0.168 | [0.165; 0.171] | 17.4% |
|  | 5 | OLS | Income support history | 0.083 | [0.080; 0.085] | 59.5% |
| Economist models | 6 | OLS | Heuristic | 0.141 | [0.138; 0.144] | 30.6% |
|  | 7 | OLS | Heuristic + Income support history | 0.077 | [0.074; 0.079] | 62.3% |
|  | 8 | OLS | Model 7 + Location | 0.077 | [0.074; 0.079] | 62.5% |
| Actuarial | 9 | OLS | Carer + Student + Parent | 0.190 | [0.188; 0.192] | 6.5% |
| Machine Learning | 10 | LASSO | All baseline inputs | 0.052 | [0.050; 0.054] | 74.7% |
|  | 11 | SVR | All baseline inputs | 0.050 | [0.048; 0.052] | 75.4% |
|  | 12 | Boosting | All baseline inputs | 0.050 | [0.048; 0.053] | 75.5% |
|  | 13 | Ensemble | All baseline inputs | 0.048 | [0.046; 0.050] | 76.3% |

*Notes*: The outcome variable is the proportion of time an individual received an income support payment in 2015-2019. Each row is a different model. MSE is the out-of-sample mean-squared error computed using the 20% hold-out sample (Step 3 as described in Online Appendix B.1). The numbers in brackets are bootstrapped 95% confidence intervals for hold-out prediction performance (Step 4 in Online Appendix B.1). $R^2$ is the squared correlation between the fitted outcome and actual outcome in the 20% hold-out sample. Model 6 inputs (measured in 2014, corresponding missing indicators added whenever necessary): sex, age, migration status, ethnicity, parent status, marital status, state of residence, education level, unemployment status. Location data in Model 7 include SA3 region data: there are 329 unique geographic regions. The actuarial Model 9 identifies three groups with the largest estimated lifetime welfare costs: young carers age 24 or under ("Carer"); students receiving a particular Income Support payment ("student"); and young parents age 18 or under ("parent"). All variables are described in Online Appendix A.

**Table 3: Top ten predictors selected by the LASSO and Boosting algorithms.**

| Order of Importance | Level of Importance | Predictors (calculated in 2014) | Variable Name |
|---|---|---|---|
| LASSO | Post-LASSO Coefficients | | |
| 1 | -1.482 | Benefit fluctuation | p_sdpy |
| 2 | 0.423 | Total benefit received | p_totpy2014 |
| 3 | 0.345 | Benefit duration (any income support payment) | p_isdur14 |
| 4 | -0.341 | Annual employment income | p_totinc2014 |
| 5 | -0.270 | Wage rate | p_wage2014 |
| 6 | 0.236 | Number of residential moves | p_totmoves |
| 7 | -0.219 | Number of sanctions | p_numsus2014 |
| 8 | 0.216 | Age: below 25 (on 1 Jan 2014) | p_aged1 |
| 9 | -0.214 | Maximum number of simultaneous jobs held | p_maxsimjob2014 |
| 10 | 0.214 | Received benefit in December | p_Dec2014 |
| Boosting | Influence Parameters | | |
| 1 | 69.538 | Benefit duration (any income support payment) | p_isdur14 |
| 2 | 11.584 | Number of jobs (missing) | p_numjob2014miss |
| 3 | 4.313 | Received benefit in quarter four | p_qr42014 |
| 4 | 3.788 | Benefit terminated for non-fulfillment of activity test | p_evsannb2014 |
| 5 | 3.267 | Benefit duration on non-main income support benefits | p_othdur2014 |
| 6 | 1.514 | Income fluctuation | p_sdinc |
| 7 | 0.605 | Age: 59-64 (on 1 Jan 2014) | p_agecatd9 |
| 8 | 0.584 | Ever on Age Pension payment | p_evage14 |
| 9 | 0.533 | Benefit duration (Disability Support Pension) | p_dspdur2014 |
| 10 | 0.320 | Annual hours | p_tothr2014 |

*Notes*: This variable list the top 10 inputs selected by LASSO (Model 10 Table 2) and Boosting (Model 12 Table 2) when predicting the proportion of time an individual received an income support payment in 2015-2019. All listed variables were measured in 2014. LASSO selected 323 variables and Boosting selected 72 variables. The missing indicator selected by Boosting – "Number of jobs (missing)" - is perfectly collinear with the variable "Ever on Income Support" (p_evis14). In LASSO, the reported coefficients are computed by taking the selected variables and estimating an OLS linear regression with them. In Boosting, the influence of an input depends on the number of times a variable is chosen across all iterations (trees) and its overall contribution to the log-likelihood function; such values are then standardised to sum up to 100. All variables are described in Online Appendix A.

**Table 4: Prediction (out-of-sample) accuracy for time on income support. Extensions.**

|  | Model | Predictors | MSE Mean | MSE C.I. | $R^2$ |
|---|---|---|---|---|---|
| 1 | OLS | Top 10 predictors from the LASSO model | 0.073 | [0.057; 0.061] | 64.1% |
| 2 | OLS | Top 10 predictors from the Boosting model | 0.056 | [0.054; 0.058] | 72.5% |
| 3 | OLS | Union of the top predictors from Models 1 and 2 | 0.056 | [0.054; 0.058] | 72.6% |
| 4 | OLS | Indigenous | 0.201 | [0.199; 0.203] | 1.3% |
| 5 | OLS | Country of birth | 0.198 | [0.196; 0.200] | 2.6% |
| 6 | OLS | Heuristic + Income support history + ZIP code | 0.080 | [0.077; 0.082] | 60.8% |
| 7 | OLS | Heuristic + Parental welfare receipt | 0.140 | [0.137; 0.143] | 31.3% |
| 8 | OLS | Heuristic + Income support history + Parental welfare receipt | 0.080 | [0.074; 0.079] | 62.6% |
| 9 | FOR | Heuristic | 0.140 | [0.137; 0.143] | 31.2% |
| 10 | Boosting | Heuristic | 0.128 | [0.124; 0.131] | 37.3% |
| 11 | LASSO | Heuristic | 0.141 | [0.138; 0.144] | 30.6% |
| 12 | LASSO | All baseline inputs (plus interactions) | 0.049 | [0.047; 0.051] | 76.0% |
| 13 | OLS | Heuristic (excluding those 100% on welfare 2011-14) | 0.119 | [0.115; 0.123] | 32.4% |
| 14 | OLS | Heuristic + Income support history (excluding those 100% on welfare 2011-14) | 0.082 | [0.079; 0.085] | 53.6% |
| 15 | LASSO | Full model (excluding those 100% on welfare 2011-14) | 0.059 | [0.056; 0.061] | 66.8% |

*Notes*: The outcome variable is the proportion of time an individual received an income support payment in 2015-2019. Each row is a different model. Location data in Model 6 include a total of 2,456 ZIP codes (or postcodes). Model 7 includes a separate set of binary variables recording the income support receipt of the primary female carer, the primary male carer, and any carer. Income support payments include: any type of income support; disability payment; unemployment benefit; carer payment; parenting payment; partner payment; age pension; financial difficulty payments. See also notes in Table 2. FOR stands for Fractional Outcome Regression (Probit). Models 13-15 exclude those who are on income support for 100 percent of time from 2011 to 2014, inclusive. The sample size in these models is 36,358. This is the 1% random sample after the deletion of those on income support for 100 percent of time between 2011 and 2014.

**Table 5: Prediction (out-of-sample) accuracy for time on Unemployment Benefits.**

|  |  | Model | Predictors | MSE Mean | MSE C.I. | $R^2$ |
|---|---|---|---|---|---|---|
| Simple models | 1 | OLS | Constant | 0.206 | [0.203; 0.209] | 0.0% |
|  | 2 | OLS | Sex | 0.204 | [0.202; 0.207] | 0.9% |
|  | 3 | OLS | Education | 0.201 | [0.198; 0.204] | 2.7% |
|  | 4 | OLS | Age | 0.083 | [0.080; 0.087] | 59.6% |
|  | 5 | OLS | Income support history | 0.125 | [0.122; 0.128] | 39.5% |
| Economist models | 6 | OLS | Heuristic | 0.036 | [0.034; 0.038] | 82.8% |
|  | 7 | OLS | Heuristic + Income support history | 0.035 | [0.034; 0.037] | 83.1% |
|  | 8 | OLS | Model 7 + Location | 0.035 | [0.033; 0.037] | 83.1% |
| Machine Learning | 9 | LASSO | All baseline inputs | 0.029 | [0.027; 0.031] | 86.2% |

*Notes*: The outcome variable is the proportion of time an individual received an unemployment benefit in 2015-2019. Each row is a different model. See also notes in Table 2. All variables are described in Online Appendix A.

**Table 6: Clustering: Key characteristics.**

| Predictors | Group 1 | Group 2 | Group 3 | Group 4 | No benefit |
|---|---|---|---|---|---|
| Simultaneous jobs worked (num) | 0.013 | 0.013 | 0.025 | 0.015 | 0.004 |
| | (0.029) | (0.03) | (0.037) | (0.039) | (0.022) |
| Immigrant | 0.252 | 0.314 | 0.194 | 0.200 | 0.368 |
| | (0.435) | (0.464) | (0.396) | (0.400) | (0.483) |
| Indigenous | 0.081 | 0.051 | 0.155 | 0.092 | 0.013 |
| | (0.274) | (0.219) | (0.362) | (0.289) | (0.112) |
| Aged below 25 | 0.116 | 0.037 | 0.355 | 0.332 | 0.070 |
| | (0.321) | (0.19) | (0.479) | (0.471) | (0.256) |
| Aged 25-29 | 0.070 | 0.036 | 0.116 | 0.166 | 0.135 |
| | (0.255) | (0.186) | (0.321) | (0.372) | (0.342) |
| Aged 30-34 | 0.085 | 0.042 | 0.093 | 0.160 | 0.256 |
| | (0.28) | (0.201) | (0.291) | (0.366) | (0.437) |
| Aged 35-39 | 0.093 | 0.081 | 0.067 | 0.075 | 0.226 |
| | (0.291) | (0.273) | (0.249) | (0.264) | (0.419) |
| Aged 40-44 | 0.081 | 0.095 | 0.069 | 0.058 | 0.154 |
| | (0.274) | (0.294) | (0.254) | (0.235) | (0.362) |
| Aged 45-49 | 0.116 | 0.099 | 0.075 | 0.045 | 0.100 |
| | (0.321) | (0.298) | (0.263) | (0.207) | (0.300) |
| Aged 50-54 | 0.120 | 0.106 | 0.086 | 0.043 | 0.046 |
| | (0.326) | (0.309) | (0.281) | (0.202) | (0.21) |
| Aged 55-59 | 0.081 | 0.122 | 0.060 | 0.049 | 0.010 |
| | (0.274) | (0.327) | (0.238) | (0.215) | (0.099) |
| Aged 60-64 | 0.105 | 0.211 | 0.078 | 0.044 | 0.003 |
| | (0.307) | (0.408) | (0.269) | (0.204) | (0.053) |
| Ever married in 2014 | 0.221 | 0.356 | 0.143 | 0.138 | 0.736 |
| | (0.416) | (0.479) | (0.35) | (0.345) | (0.441) |
| Carer payment | 0.194 | 0.108 | 0.029 | 0.104 | 0.000 |
| | (0.396) | (0.31) | (0.169) | (0.305) | (0.000) |
| Female | 0.419 | 0.574 | 0.468 | 0.704 | 0.657 |
| | (0.494) | (0.495) | (0.499) | (0.457) | (0.475) |
| Parenting payments | 0.004 | 0.077 | 0.084 | 0.424 | 0.000 |
| | (0.062) | (0.266) | (0.278) | (0.494) | (0.000) |
| Observations | 258 | 3,108 | 1,126 | 1,009 | 712 |

*Notes*: Summary statistics (mean and standard deviation) are reported for each group identified by the hierarchical clustering algorithm. Individuals who were identified as at-risk of long-term welfare receipt have been divided into four groups (first four column of results). The last column reports summary statistics for the group of individuals predicted to have no future receipt. All variables have been rescaled to be between 0 and 1. Full set of variables reported in Table B8. All variables are described in Online Appendix A.

Online Appendix for

# Using Machine Learning to Create an Early Warning System for Welfare Recipients

**Appendix A. Variable description**

**A.1 Dependent (outcome) variables**

*Proportion of time spent on income support from 2015 to 2018 (isprop).* This is the total number of days an individual received income support payments from 1 January 2015 to 31 December 2018 divided by the total number of days spanning this period. Income support payments include the following:

- Age Pension
- Austudy
- Bereavement Allowance
- Carer Payment
- Disability Support Pension
- Exceptional Circumstances Payment
- Farm Family Restart
- Mature Age Allowance
- Mature Age Partner Allowance
- Newstart Mature Age Allowance
- Newstart Allowance
- Parenting Payment Partnered
- Parenting Payment Single
- Partner Allowance
- Sickness Allowance
- Special Benefit
- Widow Allowance
- Wife Pension Age
- Wife Pension DSP
- Widow B Pension
- Youth Allowance (Apprentice)
- Youth Allowance (Other)
- Youth Allowance (Student)

- Youth Training Allowance

**A.2 Independent (input) variables**

For each variable, missing values (if any) have been set to zero and a new binary variable has been generated to indicate the observations that are missing.

*Demographics*

*Female (p_female)* records whether an individual is recorded as a woman.

*Immigrant (p_immi)* records whether an individual is an immigrant, i.e. not born in Australia.

*Indigenous (p_indig)* records whether an individual is indigenous, which includes being an Aboriginal or Torres Strait Islander.

*Age* records an individual's age on 1 January 2014. We have deleted anyone aged less than 15 and older than 66 as on 1 January 2014. We have created a series of binary indicators for ages 15 to 66 *(p_aged\*)* as well as for the following age categories:

- Below 25 *(p_agecatd1)*
- 25-29 and then in increments of 5 years up to age 64 *(p_agecatd2- p_agecatd9)*
- 65 and above *((p_age65ab)*

*Country of birth* records whether or not an individual was born in:

- Australia *(p_auborn)*
- World regions including:
    - Asia *(p_Regiond1)*
    - Middle East and North Africa *(p_Regiond2)*
    - Europe *(p_Regiond3)*
    - North America *(p_Regiond4)*
    - Central America and the Caribbean *(p_Regiond5)*
    - South America *(p_Regiond6)*
    - Sub-Saharan Africa *(p_Regiond7)*
    - Oceania and Pacific Islands *(p_Regiond8)*
    - Unclassified or Unknown *(p_Regiond9)*

- The top source countries for Australian immigrants including:
    - South Africa *(p_saborn)*
    - United States of America *(p_usborn)*
    - United Kingdom *(p_ukborn)*
    - China, excluding Hong Kong *(p_cnborn)*
    - India *(p_inborn)*
    - Hong Kong *(p_hkborn)*
    - Taiwan *(p_twborn)*
    - Philippines *(p_phborn)*
    - Indonesia *(p_indoborn)*
    - Fiji *(p_fiborn)*
    - Vietnam *(p_vtborn)*
- An English-speaking country *(p_esc)* including:
    - Papua New Guinea
    - Tuvalu
    - New Zealand
    - South Africa
    - United States of America
    - United Kingdom
    - Canada
    - Fiji
    - Tokelau
    - Tonga
    - Vanuatu
    - Samoa

*Income support and government benefit receipt history*

*IS history* in 2014 records whether an individual had ever received any type of income support payments during 2014 *(p_evis14)* as well as receipt for the following (types of) income support payments:

- Disability Support Pension *(p_evdsp14)*

- Carer Payment *(p_evcar14)*
- Age Pension *(p_evage14)*
- Unemployment benefits *(p_evune14)* including Newstart Mature Age Allowance, Newstart Allowance, Youth Allowance (Other) and Youth Training Allowance. This variable is used to measure unemployment history during 2014
- Parenting payments *(p_evpar14)* including Parenting Payment Partnered and Parenting Payment Single
- Partner payments *(p_evpart14)* including Mature Age Partner Allowance, Partner Allowance, Wife Pension Age and Wife Pension DSP
- Financial difficulty payments *(p_evcri2014)* including Exceptional Circumstances Payment and Special Benefit

*Receipt of Remote Area Allowance in 2014 (p_evraa2014)* records whether an individual had ever received Remote Area Allowance in 2014. Remote Area Allowance is not an income support payment in its own right, but rather an additional payment that income support recipients can receive if they live in a remote area.

*Receipt of one-off Crisis Payment in 2014* (*p_cripay2014*) records whether an individual had ever received a one-off Crisis Payment in 2014. Crisis Payment is a one-off payment that income support recipients can access if they are in severe financial difficulty due to extreme circumstances.

*Proportion of time spent on income support in 2014 (p_isprop14)*. This is the total number of days an individual received income support payments from 1 January 2014 to 31 December 2014 divided by the total number of days spanning this period.

*Income support churning (p_ischurn)* records whether an individual had ever left and came back onto income support in 2014.

*Income support transfer (p_istransfer)* records whether an individual had ever transferred between different income support payments in 2014.

*Times transferred between different income support payments (p_numistran)* records the number of times an individual had transferred between different income support payments in 2014.

*Seasonality in Centrelink administered payment receipt.* A set of indicators recording whether an individual had ever received a Centrelink administered payment for each month *(p_Jan2014-p_Dec2014)*, as well as quarter of 2014 *(p_qr12014-p_qr42014)*.

*Duration on benefits in 2014 (p_bendur2014)* measures the total number of days an individual was in the Centrelink system in 2014 including the days they were receiving a government payment (including income support payments, Family Tax Benefits and other payments such as rental assistance) and the days when they were suspended from a payment.

*Government benefit durations in 2014* record the total number of days an individual was receiving income support payments *(p_isdur14)*, Family Tax Benefits *(p_ftbdur2014)* and the following (types of) income support payments in 2014:

- Disability Support Pension *(p_dspdur2014)*
- Carer Payment *(p_cardur2014)*
- Age Pension *(p_agedur2014)*
- Unemployment benefits *(p_ubdur2014)* including Newstart Mature Age Allowance, Newstart Allowance, Youth Allowance (Other) and Youth Training Allowance. This is used to measure unemployment history during 2014.
- Parenting payments (*p_pardur2014*) including Parenting Payment Partnered and Parenting Payment Single
- Partner payments *(p_partdur2014)* including Partner Allowance, Wife Pension Age, Wife Pension DSP and Mature Age Partner Allowance
- Crisis payments *(p_cridur2014)* including Exceptional Circumstances Payment and Special Benefit
- Other income support payments *(p_othdur2014)* including Austudy, Bereavement Allowance, Farm Family Restart, Widow Allowance and Widow B Pension

*Total amount of government benefit received in 2014 (p_totpy)* records the total amount of government benefits received by an individual in 2014. These include regular payments, once-off payments, advanced payments and payments to third parties on behalf of an individual in 2014.

*Government benefit fluctuation in 2014 (p_sdpy)* records the fluctuation in the amount of government benefit received by an individual in 2014 as measured by the standard deviation in the amount of government benefit received on a bi-weekly basis.

*Ever received advanced payment in 2014 (p_evadv2014)* records whether an individual had ever received advanced payments in 2014. This is a companion variable to *Ever received a one-off Crisis Payment in 2014* (*p_cripay2014*). The difference between p_evadv2014 and p_cripay2014 is that the former shifts forward a scheduled payment whereas the latter is an additional payment.

*Income support neighbourhood effects*

*Maximum proportion of income support recipients in neighbourhood in 2014 (p_maxisnei2014)* measures the maximum percentage of income support recipients in any of the postcodes or SA1 regions an individual has lived in 2014.

*Sanctions*

*Sanctioned in 2014 (p_evsus2014)* records whether an individual had ever had their Centrelink administered payment suspended during 2014.

*Number of sanctions in 2014 (p_numsus2014)* records the number of times an individual had had their Centrelink administered payment suspended during 2014.

*Government Benefit terminated for non-fulfillment/breach in 2014 (p_evsannb2014)* record whether an individual had ever had a government benefit terminated for non-fulfillment or breach of activity agreements to income support receipt in 2014.

*Personal relationships*

*Ever single in 2014 (p_evsing2014)* records whether an individual had ever been single (having a marital status that is either single, separated, divorced or widowed) in 2014 when they were in the Centrelink system, i.e. they were receiving a Centrelink administered payment (including income support payments, Family Tax Benefits and other payments such as rental assistance) or were suspended from a payment.

*Ever married in 2014 (p_evmar2014)* records whether an individual had ever been married in 2014 when they were in the Centrelink system, i.e. they were receiving a Centrelink administered payment (including income support payments, Family Tax Benefits and other payments such as rental assistance) or were suspended from a payment.

*Ever in a de facto partnership in 2014 (p_evdef2014)* records whether an individual had ever been in a de facto partnership in 2014 when they were in the Centrelink system, i.e. they were receiving a Centrelink administered payment (including income support payments, Family Tax Benefits and other payments such as rental assistance) or were suspended from a payment.

*Ever separated in 2014 (p_evsep2014)* records whether an individual had ever been separated in 2014 when they were in the Centrelink system, i.e. they were receiving a Centrelink administered payment (including income support payments, Family Tax Benefits and other payments such as rental assistance) or were suspended from a payment.

*Ever divorced in 2014 (p_evdiv2014)* records whether an individual had ever been divorced in 2014 when they were in the Centrelink system, i.e. they were receiving a Centrelink administered payment (including income support payments, Family Tax Benefits and other payments such as rental assistance) or were suspended from a payment.

*Ever widowed in 2014 (p_evwid2014)* records whether an individual had ever been widowed in 2014 when they were in the Centrelink system, i.e. they were receiving a Centrelink administered payment (including income support payments, Family Tax Benefits and other payments such as rental assistance) or were suspended from a payment.

*Ever changed relationship status in 2014 (p_evchms2014)* records whether an individual had ever changed relationship status between being single (single, separated, divorced or widowed) and partnered (de facto or married) in 2014 when they were in the Centrelink system, i.e. they were receiving a Centrelink administered payment (including income support payments, Family Tax Benefits and other payments such as rental assistance) or were suspended from a payment.

*Number of times changed relationship status in 2014 (p_numchms2014)* records the number of times an individual had changed relationship status between being single (single, separated, divorced or widowed) and partnered (de facto or married) in 2014 when they were in the Centrelink system, i.e. they were receiving a Centrelink administered payment (including income support payments, Family Tax Benefits and other payments such as rental assistance) or were suspended from a payment.

*Relationship duration in 2014 (p_reldur2014)* records the number of days from 1 January 2014 to 31 December 2014 for which an individual had known relationship status in the DOMINO data, i.e. either single, separated, divorced, widowed, de facto partnered or married.

*Single parent in 2014 (p_evlp2014)* records whether an individual had ever been a single parent in 2014. We have constructed this variable from first identifying whether an individual had ever been single and whether an individual had even been a parent within each half-month period of 2014, separately. If an individual had been both single at a point of time during any of these 24-period and the same individual had also been a parent within the same period, we have coded the individual as having been a single parent in 2014.

### Children

*With child in 2014 (p_parent2014)* records whether an individual had ever had a child in their care in 2014 when they in the Centrelink system, i.e. they were receiving a Centrelink administered payment (including income support payments, Family Tax Benefits and other payments such as rental assistance) or were suspended from a payment.

*Children below 16 in 2014 (p_kidb16in2014)* record whether an individual had ever had a child that was under the age of 16 in their care in 2014 when they in the Centrelink system, i.e. they were receiving a Centrelink administered payment (including income support payments, Family Tax Benefits and other payments such as rental assistance) or were suspended from a payment.

*Children below 10 in 2014 (p_kidb10in2014)* record whether an individual had ever had a child that was under the age of 10 in their care in 2014 when they in the Centrelink system, i.e. they were receiving a Centrelink administered payment (including income support payments, Family Tax Benefits and other payments such as rental assistance) or were suspended from a payment.

*Children below 5 in 2014 (p_kidb5in2014)* record whether an individual had ever had a child that was under the age of 5 in their care in 2014 when they were in the Centrelink system, i.e. they were receiving a Centrelink administered payment (including income support payments, Family Tax Benefits and other payments such as rental assistance) or were suspended from a payment.

*Age of youngest in 2014 (p_chageyng2014)* records the age of the youngest child in an individual's care in 2014.

*Age of eldest in 2014 (p_chageeld2014)* records the age of the eldest child in an individual's care in 2014.

*Average age of children in 2014 (p_chagemean2014)* records the average age of the children in an individual's care in 2014.

*Number of children in 2014 (p_numch2014)* records the number of children (aged below 24) under an individual's care in 2014.

*Number of children by age group* records the number of children an individual had under their care in 2014 that fell into the following age categories:

- Below 1 (*p_numchb1*)
- At least 1, below 3 (*p_numch1t3*)
- At least 3, below 5 (*p_numch3t5*)
- At least 5, below 8 (*p_numch5t8*)
- At least 8, below 10 (*p_numch8t10*)
- At least 10, below 15 (*p_numch10t15*)
- 15 and above (*p_numch15ab*)

*Presence of children by age (p_havekidaged*)* is a set of binary indicators for whether an individual had a child under their care in 2014 that was aged for each year from 0 to 16, respectively.

*Presence of children by age group* is a set of binary indicators for whether an individual had a child under their care in 2014 that fell into the following age categories:

- Below 1 (*p_havekidb1*)
- At least 1, below 3 (*p_havekid1t3*)
- At least 3, below 5 (*p_havekid3t5*)
- At least 5, below 8 (*p_havekid5t8*)
- At least 8, below 10 (*p_havekid8t10*)
- At least 10, below 15 (*p_havekid10t15*)
- 15 and above (*p_havekid15ab*)

*Employment*

*Annual employment income in 2014 (p_totinc2014)* records the total dollar amount an individual earned through employment in 2014 when the individual was in the income support system, i.e. they were receiving income support payments or were suspended from a payment. We have not top-coded this variable, but we have created a new variable to indicate values at the 99$^{th}$ percentile or above of the entire income distribution for the sample *(p_inctc)*.

*Annual hours in 2014 (p_tothr2014)* records the total hours worked by an individual in 2014 when the individual was in the income support system, i.e. they were receiving income support payments or were suspended from a payment. We have top-coded this variable so that its maximum is 5200 hours and we have created a new variable to indicate values that have been top-coded *(p_thrtc)*.

*Wage rate in 2014 (p_wage2014)* records the wage rate calculated by annual employment income in 2014 over annual hours in 2014.

*Average daily hours worked in 2014 (p_avhr2014)* records the average daily hours worked by an individual in 2014 across income spells when the individual was in the income support system, i.e. they were receiving income support payments or were suspended from a payment.

*Minimum daily employment income in 2014 (p_mininc2014)* records the minimum daily income an individual earned through employment in 2014 when the individual was in the income support system, i.e. they were receiving income support payments or were suspended from a payment. We have not top-coded this variable, but we have created a new variable to indicate values at the 99$^{th}$ percentile or above of the entire income distribution for the sample *(p_dinctc)*.

*Minimum daily hours in 2014 (p_minhr2014)* records the minimum daily hours worked by an individual in 2014 when the individual was in the income support system, i.e. they were receiving income support payments or were suspended from a payment.

*Zero hour contracts in 2014 (p_evzh)* records whether an individual had ever worked in a zero-hour contract (where they reported zero income and zero hours, which is distinguished from the case of positive income and zero hours which reflect bonuses received for a given job) for continuous income in 2014 when the individual was in the income support system, i.e. they were receiving income support payments or were suspended from a payment. If the individual was never in the income support system in 2014 but had worked in a zero-hour contract while receiving some non-income support government benefits in 2014, then we have also set this variable to 1 for this person but otherwise this variable is missing for those who never received any income support in 2014. This is a proxy for underemployment.

*Number of jobs in 2014 (p_numjob2014)* records the total number of jobs, as measured by a change in employers, an individual had worked in when the individual was in the income support system, i.e. they were receiving income support payments or were suspended from a payment.

*Maximum number of simultaneous jobs worked in 2014 (p_maxsimjob2014)* records the maximum number of simultaneous jobs, as measured by working with different employers at the same time, an individual had worked in when the individual was in the income support system, i.e. they were receiving income support payments or were suspended from a payment.

*Number of substantial employment spells (p_lempnum2014)* records the number of substantial employment spells an individual had in 2014. To measure substantial employment spells, we ignore any breaks in employment of 28 days or less.

*Minimum wage worker in 2014* (*p_minwage2014*) records whether an individual had worked in a job whose hourly wage was at or below the federal minimum wage in July 2014 when the individual was in the income support system, i.e. they were receiving income support payments or were suspended from a payment.

*Work instability*

*Income fluctuation in 2014 (p_sdinc)* records the fluctuation in the amount of an individual's employment income in 2014 when the individual was in the income support system, i.e. they were receiving income support payments or were suspended from a payment, as measured by the standard deviation in the amount of employment income earned on a bi-weekly basis.

*Hours fluctuation in 2014 (p_sdhr2014)* records the fluctuation in the amount of an individual's employment hours in 2014 when the individual was in the income support system, i.e. they were receiving income support payments or were suspended from a payment, as measured by the standard deviation in hours worked on a bi-weekly basis. Note that we have first top-coded fortnightly total working hours to 200 hours.

*Average length of employment tenure (p_maxjobten2014)* records the average length of tenure with a given employer in 2014 when an individual was in the income support system, i.e. they were receiving income support payments or were suspended from a payment.

*Maximum length of employment tenure (p_avjobten2014)* records the maximum length of tenure with a given employer in 2014 when an individual was in the income support system, i.e. they were receiving income support payments or were suspended from a payment.

*Number of times that working hours changed in 2014 (p_numhrch)* records the number of times an individual changed their working hours by more than 20% on a bi-weekly basis when they were in the income support system, i.e. they were receiving income support payments or were suspended from a payment.

*Intensity of working hours change in 2014* records whether an individual had changed working hours on a bi-weekly basis when they were in the income support system, i.e. they were receiving income support payments or were suspended from a payment. We distinguish magnitudes of change by the following categories:

- Decreased by 51% or more (*p_evhrd50m)*
- Decreased by 31-50% (*p_evhrd31t50)*
- Decreased by 11-30% (*p_evhrd11t30)*
- Remained relatively unchanged: hours fluctuate by up to 10% (*p_evsamehr)*
- Increased by at least 11-30% (*p_evhri11t30)*
- Increased by at least 31-50% (*p_evhri31t50)*
- Increased by 51% or more (*p_evhri50m)*

*Ever had increased work instability in 2014 (p_evacqvar2014)* records whether an individual had ever had increased work instability in 2014 when they were in the income support system, i.e. they were receiving income support payments or were suspended from a payment, as measured by whether they had only continuous income spells in 2014 and then 'acquired' new spells of variable income. If the individual was never in the income support system in 2014 but had 'acquired' new spells of variable income while receiving some non-income support government benefits in 2014, then we also set this variable to 1 for this person but otherwise this variable is set to missing for those who never received any income support in 2014

*Ever had increased work instability within employer in 2014 (p_morewithinwi2014)* records whether an individual had ever stayed with same employer but went from having continuous to variable income with the employer in 2014 when they were in the income support system, i.e. they were receiving income support payments or were suspended from a payment. If an individual was never in the income support system in 2014 but had increased work instability within the same employer in 2014, then we also set this variable to 1 for this person but otherwise this variable is set to missing for those who never received any income support in 2014.

*Ever had changed days worked in the week in 2014 (p_evchwday2014)* records whether an individual had ever changed days worked in the week in 2014 when they were in the income support system, i.e. they were receiving income support payments or were suspended from a payment. This is derived from examining the day of the week on which individuals' income spell start and end on. For example, if their reporting changed from being from Sunday to Saturday to being from Monday to Sunday instead, we assume they had changed the days in the week that they worked on in 2014. If the individual was never in the income support system in 2014 but had changed days worked in the week in 2014, then we also set this variable to 1 for this person but otherwise this variable is set to missing for those who never received any income support in 2014.

*Housing*

*Homeless in 2014 (p_hl2014)* records whether an individual had ever been homeless during 2014 when they were in the Centrelink system, i.e. they were receiving a Centrelink administered payment (including income support payments, Family Tax Benefits and other payments such as rental assistance) or were suspended from a payment.

*Homeowner in 2014 (p_homeowner2014)* records whether an individual had ever been a homeowner in 2014 when the individual was in the Centrelink system, i.e. they were receiving a Centrelink administered payment (including income support payments, Family Tax Benefits and other payments such as rental assistance) or were suspended from a payment. The definition of home ownership encompasses the following:

- Jointly/singly owns or purchasing home
- Owns jointly with their partner
- Owns jointly with someone other than partner

- Owns home but can't be classified by above

*Rent type in 2014* records whether an individual had ever paid the following types of rent in 2014 when they were in the Centrelink system, i.e. they were receiving a Centrelink administered payment (including income support payments, Family Tax Benefits and other payments such as rental assistance) or were suspended from a payment:

- Public *(p_pubrent2014)*
- Private *(p_prirent2014)*
- Other *(p_othrent2014)*

*Accommodation type in 2014* records whether an individual had ever lived in the following types of accommodation in 2014 when they were in the Centrelink system, i.e. they were receiving a Centrelink administered payment (including income support payments, Family Tax Benefits and other payments such as rental assistance) or were suspended from a payment:

- Shared accommodation *(p_shraccom2014)*
- Non-shared accommodation *(p_nshraccom2014)*
- Lived with their parents *(p_parenthome2014)*
- Exempt *(p_exemaccom2014)*
- Accommodation type unknown (*p_unkaccom2014)*

*Accommodation duration in 2014 (p_accomdur2014)* records the number of days from 1 January 2014 to 31 December 2014 for which an individual has known accommodation information in the DOMINO data.

*Rent paid in 2014 (p_totrent2014)* records the total amount of rent paid by an individual in 2014 when they were in the Centrelink system, i.e. they were receiving a Centrelink administered payment (including income support payments, Family Tax Benefits and other payments such as rental assistance) or were suspended from a payment.

*Rental Assistance Receipt in 2014 (evra2014)* record whether an individual had ever received Rental Assistance in 2014 including the following:

- Rental Assistance Family
- Rental Assistance Parenting

- Rental Assistance Newstart
- Rental Assistance Pension
- Rental Assistance Abstudy

*Rental burden in 2014 (p_rentstr2014)* records the rental burden an individual experienced in 2014 when they were in the Centrelink system, i.e. they were receiving a Centrelink administered payment (including income support payments, Family Tax Benefits and other payments such as rental assistance) or were suspended from a payment. It is calculated as the total rent paid in 2014 divided by the sum of the total amount of government benefits received in 2014 and the annual employment income in 2014. We have not top-coded this variable, but we have created an indicator for rental burden being greater than 1 *(p_rstc)*.

*Location and Residential movement*

In the DOMINO data, we have postcode (ZIP code) information. Also, where geocoding can be applied we use geographic information pertaining to the Australian Statistical Geography Standard (ASGS), a framework used by the Australian Bureau of Statistics that covers the all of Australia and allow a variety of different geographical aggregation for which data analysis can be undertaken at. The ASGS has 4 levels.

- Statistical Areas Level 1 (SA1s) are the smallest geographic aggregation, with a population between 200 and 800 people
- Statistical Areas Level 2 (SA2s) are aggregations of SA1s and are meant to reflect functional areas that represent a community that interacts together socially and economically. Its population size ranges from 3000 to 25,000 persons
- Statistical Areas Level 3 (SA3s) are aggregations of SA2s. They are broad geographic regions that are about the size of a town or a cluster of suburbs around a major business and transport centre. Population size ranges from 30,000 to 130,000

For the DOMINO, we have SA1 and SA2 codes, from which SA3 codes can be easily derived. For the main set of geographic variables, we use the SA3 level.

*Moved in 2014 (p_evmove)* records whether an individual had ever moved during 2014 when they were in the Centrelink system, i.e. they were receiving a Centrelink administered payment (including income support payments, Family Tax Benefits and other payments such as rental assistance) or were suspended from said payment. We proxy residential moves with movement between SA1 regions or postcodes.

*Number of residential moves in 2014 (p_totmoves)* measures the number of times an individual had moved across different SA1 regions or postcodes in 2014.

*Location in 2014* include a set of indicators recording whether an individual had ever lived in each of the eight states in Australia in 2014 (p_evnsw, p_evvic, p_evwa, p_evsa, p_evact, p_evnt, p_evtas, p_evqld), as well as each of the SA3 regions in Australia in 2014 *(p_evlsa3_*)*.

*Residential duration in 2014 (p_resdur2014)* measures the total number of days an individual's location details are recorded2014.

*SEIFA movements in 2014* is a set of three indicators for whether an individual had moved up *(p_upseifa2014)* or down at least a decile *(p_downseifa2014)* on the Socioeconomic Index of Disadvantage or stayed within the same SEIFA decile *(p_sameseifa2014)* contingent on the individual having moved across postcodes or SA1 regions. If an individual has not moved in 2014, we set these three indicators to 0.

*Temporary housing in 2014 (p_evtempres2014)* records whether an individual had ever lived in temporary housing in 2014 when they were in the Centrelink system, i.e. they were receiving a Centrelink administered payment (including income support payments, Family Tax Benefits and other payments such as rental assistance) or were suspended from a payment. This is measured by an individual having a residence spell that does not coincide in timing with a permanent residence spell in 2014 because this indicates that the individual was using a Postal Office Box for correspondences rather than living in temporary housing in 2014.

*Education*

*Ever studied in 2014 (p_evsd2014)* records whether an individual had ever studied when they were in the Centrelink system, i.e. they were receiving a Centrelink administered payment (including income support payments, Family Tax Benefits and other payments such as rental assistance) or were suspended from said payment in 2014.

*Highest education attainment in 2014* records an individual's education attainment up to the last point they were observed in 2014 for the following levels:

- Year 10 or below *(p_eduy10)*
- Year 12 or below *(p_eduy12)*
- Certificate or below *(p_educert)*
- Certificate 1 or below *(p_educert1)*
- Certificate 2 or below *(p_educert2)*
- Certificate 3 or below *(p_educert3)*
- Certificate 4 or below *(p_educert4)*
- Diploma or below *(p_edudip)*
- Bachelor degree or below *(p_edubach)*
- Education missing *(p_edumiss)*

*Caring Responsibilities*

*Adult Carer in 2014 (p_adcare2014)* records whether an individual had ever cared for an adult when they were receiving Carer Payment in 2014.

*Children Carer in 2014 (p_chcare2014)* records whether an individual had ever cared for a child when they were receiving Carer Payment in 2014.

**A.3 Additional input variables**

Note: these inputs have been used in extension analyses (Table 4) and not in the main ML algorithms (Table 2)

*Location*

*Postcode in 2014 (p_evlpc\*)* includes a set of indicators for whether an individual had ever lived in each of the postcodes in Australia in 2014.

*Parental income support history*

*Father's income support history in 2014* records whether any of the fathers who had ever cared for an individual before 2014 had ever received any type of income support payments during 2014 *(p_dais14)*, as well as receipt for the following (types of) income support payments:

- Disability Support Pension *(p_dadsp14)*
- Carer Payment *(p_dacar14)*

- Age Pension *(p_daage14)*

- Unemployment benefits *(p_daub14)* including Newstart Mature Age Allowance, Newstart Allowance, Youth Allowance (Other) and Youth Training Allowance.

- Parenting payments *(p_dapar14)* including Parenting Payment Partnered and Parenting Payment Single

- Partner payments *(p_dapart14)* including Mature Age Partner Allowance, Partner Allowance, Wife Pension Age and Wife Pension DSP

- Financial difficulty payments *(p_dacri14)* including Exceptional Circumstances Payment and Special Benefit

*Mother's income support history in 2014* records whether any of the mothers who had ever cared for an individual before 2014 had ever received any type of income support payments during 2014 *(p_mais14)*, as well as receipt for the following (types of) income support payments:

- Disability Support Pension *(p_madsp14)*
- Carer Payment *(p_macar14)*
- Age Pension *(p_maage14)*
- Unemployment benefits *(p_maub14)* including Newstart Mature Age Allowance, Newstart Allowance, Youth Allowance (Other) and Youth Training Allowance.
- Parenting payments *(p_mapar14)* including Parenting Payment Partnered and Parenting Payment Single
- Partner payments *(p_mapart14)* including Mature Age Partner Allowance, Partner Allowance, Wife Pension Age and Wife Pension DSP
- Financial difficulty payments *(p_macri14)* including Exceptional Circumstances Payment and Special Benefit

*Parent's income support history in 2014* records whether any of the parents (fathers or mothers) who had ever cared for an individual before 2014 had ever received any type of income support payments during 2014 *(p_paris14)*, as well as receipt for the following (types of) income support payments:

- Disability Support Pension *(p_pardsp14)*
- Carer Payment *(p_parcar14)*

- Age Pension *(p_parage14)*
- Unemployment benefits *(p_parub14)* including Newstart Mature Age Allowance, Newstart Allowance, Youth Allowance (Other) and Youth Training Allowance.
- Parenting payments *(p_parpar14)* including Parenting Payment Partnered and Parenting Payment Single
- Partner payments *(p_parpart14)* including Mature Age Partner Allowance, Partner Allowance, Wife Pension Age and Wife Pension DSP
- Financial difficulty payments *(p_parcri14)* including Exceptional Circumstances Payment and Special Benefit

*Father's income support history from 2000 to 2014* records whether any of the fathers who had ever cared for an individual before 2014 had ever received any type of income support payments at any time during 2000 to 2014 *(p_dais0014)*, as well as receipt for the following (types of) income support payments:

- Disability Support Pension *(p_dadsp0014)*
- Carer Payment *(p_dacar0014)*
- Age Pension *(p_daage0014)*
- Unemployment benefits *(p_daub0014)* including Newstart Mature Age Allowance, Newstart Allowance, Youth Allowance (Other) and Youth Training Allowance.
- Parenting payments *(p_dapar0014)* including Parenting Payment Partnered and Parenting Payment Single
- Partner payments *(p_dapart0014)* including Mature Age Partner Allowance, Partner Allowance, Wife Pension Age and Wife Pension DSP
- Financial difficulty payments *(p_dacri0014)* including Exceptional Circumstances Payment and Special Benefit

*Mother's income support history from 2000 to 2014* records whether any of the mothers who had ever cared for an individual before 2014 had ever received any type of income support payments at any time during 2000 to 2014 *(p_mais0014)*, as well as receipt for the following (types of) income support payments:

- Disability Support Pension *(p_madsp0014)*

- Carer Payment *(p_macar0014)*
- Age Pension *(p_maage0014)*
- Unemployment benefits *(p_maub0014)* including Newstart Mature Age Allowance, Newstart Allowance, Youth Allowance (Other) and Youth Training Allowance.
- Parenting payments *(p_mapar0014)* including Parenting Payment Partnered and Parenting Payment Single
- Partner payments *(p_mapart0014)* including Mature Age Partner Allowance, Partner Allowance, Wife Pension Age and Wife Pension DSP
- Financial difficulty payments *(p_macri0014)* including Exceptional Circumstances Payment and Special Benefit

*Parent's income support history from 2000 to 2014* records whether any of the parents (fathers or mothers) who had ever cared for an individual before 2014 had ever received any type of income support payments at any time during 2000 to 2014 *(p_paris0014)*, as well as receipt for the following (types of) income support payments:

- Disability Support Pension *(p_pardsp0014)*
- Carer Payment *(p_parcar0014)*
- Age Pension *(p_parage0014)*
- Unemployment benefits *(p_parub0014)* including Newstart Mature Age Allowance, Newstart Allowance, Youth Allowance (Other) and Youth Training Allowance.
- Parenting payments *(p_parpar0014)* including Parenting Payment Partnered and Parenting Payment Single
- Partner payments *(p_parpart0014)* including Mature Age Partner Allowance, Partner Allowance, Wife Pension Age and Wife Pension DSP
- Financial difficulty payments *(p_parcri0014)* including Exceptional Circumstances Payment and Special Benefit

**Appendix B. ML technical details**

**B.1 ML calibration procedure**

We have followed the recommendation of Mullainathan and Spiess (2017) and split the data into two sub-samples. We have used a training sample (80% of the data) to calibrate and estimate the algorithm under each of the ML methods. We have then reported out-of-sample performance using the hold-out sample (the remaining 20% of the data). We have calibrated all ML algorithms through 5-fold cross-validation, i.e. using the following procedure:

1. Divide the 80% training sample in 5 folds.
    a. Select a possible numeric value for each tuning parameter.
        i. Train the algorithm on 4 of the 5 folds using the selected parameter values.
        ii. Predict the outcome variable for the individuals in the remaining fold and compute the mean squared error (MSE).
        iii. Repeat the above procedure five times, one for each fold.
        iv. Compute the average performance in the 5 folds.
    b. Repeat for each possible combination of the parameter values.
    c. Select the combination of the parameter values that minimises the MSE. These in-sample statistics are reported in Tables B4-B5.
2. Train the algorithm using the full 80% training sample with the selected parameter values. The resulting MSE and the square of the Pearson correlation coefficient ($R^2$) are reported in Tables B4-B5.
3. Predict the outcome variable for the individuals in the 20% hold-out sample and compute the out-of-sample MSE and the square of the Pearson correlation coefficient ($R^2$). These are the out-of-sample statistics reported in Tables 2 and 4.
4. Compute 95% confidence intervals for the out-of-sample MSE using bootstrapping. These confidence intervals are reported in Tables 2 and 4 as well.

It is important to mention that the bootstrapped confidence intervals are not the standard confidence intervals computed in regression models. As emphasised in Mullainathan and Spiess (2017), "these uncertainty estimates represent only variation of the hold-out sample for this fixed set of prediction functions, and not the variation of the functions themselves". Indeed, it is currently impossible for several ML methods to construct standard confidence intervals that are valid, even if only asymptotically (Athey and Imbens, 2019).

Since the OLS models do not require any tuning, we have simply followed steps 2-4.

**B.2 LASSO calibration**

The first method we have used in our analysis is LASSO (Least Absolute Shrinkage and Selection Operator). This method adds a penalisation term to the OLS objective function:

$$\hat{\beta}(\lambda) = \underset{\beta \epsilon \mathbb{R}^k}{\mathrm{argmin}} \sum_{i=1}^{n}(y_i - x_i'\beta)^2 + \lambda \sum_{j=1}^{k}|\beta_j|$$

Where $k$ is the number of potential predictors (potentially larger than the number of observations, $n$), and $\lambda$ is the regularization (or penalization) term. In practice, this algorithm selects variables with high predictive power, and shrinks their coefficients, while constraining all other variables to have coefficients of zero. Therefore, the key underlying assumption of LASSO is sparsity, i.e. that only a limited set of inputs in $x_i'$ play an important role in predicting the outcome variable $y_i$. LASSO is increasingly being used by economists (Mullainathan and Spiess 2017), with several applications also in studies estimating causal effects (Belloni et al. 2014, Knaus et al. 2020).

We have estimated the algorithm in Stata 16 using the built-in command *lasso linear*. We have chosen the penalisation term λ via 5-fold cross-validation in the 80% training sample using a grid search over 100 values. As shown in Table B1, the selected λ is in the interior of the interval considered for the grid search. In addition, Figure B1 plots the relation between λ and the 5-fold average cross-validation MSE (plotted using the command *cvplot* in Stata 16). This figure shows that the choice of λ substantially affects the LASSO performance, and that the selected value is indeed the one that minimises the MSE. Furthermore, as shown in Tables 2 and B4, the in-sample and out-of-sample MSE are close to each other, thus confirming that the algorithm is not overfitting or underfitting the training data.

**Table B1: LASSO tuning.**

| Outcome | λ | | | | #Predictors |
|---|---|---|---|---|---|
| | Min | Max | #Values | Selected | |
| Income support | 0.0000340 | 0.3400355 | 100 | 0.0008824 | 323 |

**Figure B1: LASSO cross-validation**

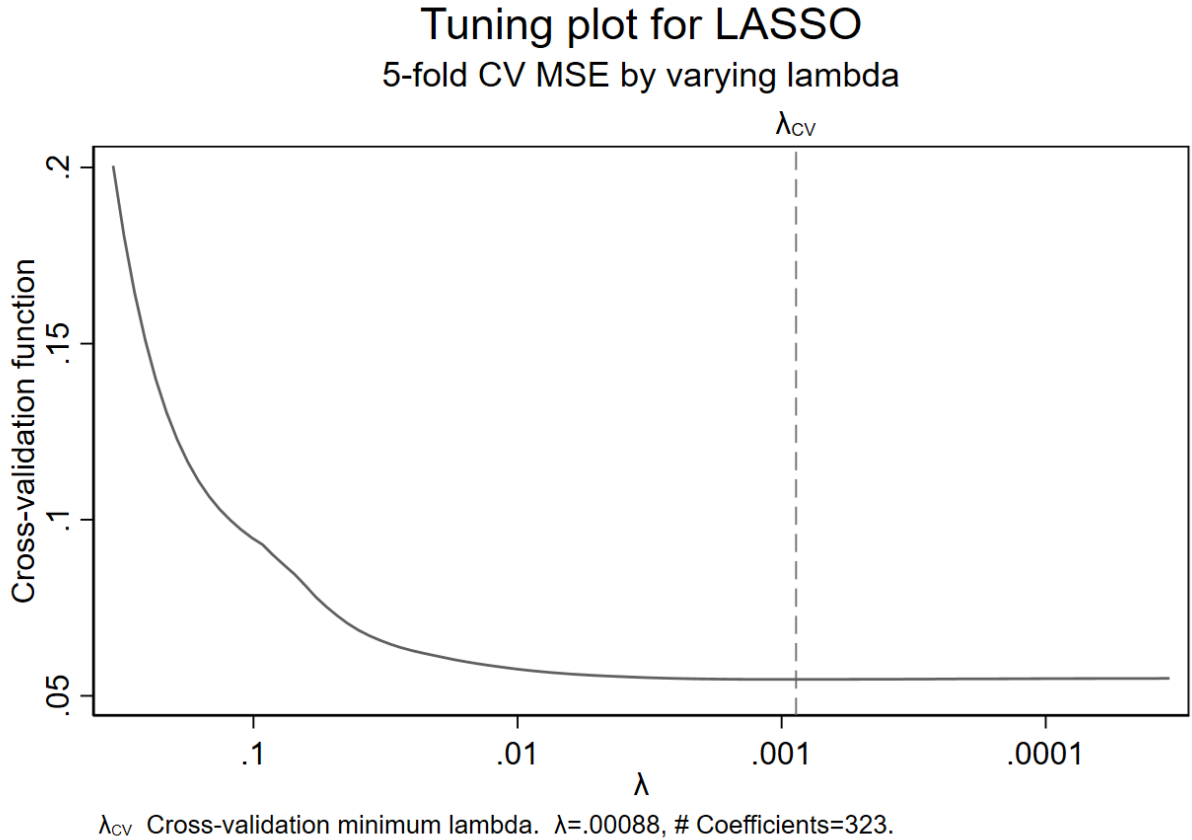

### B.3 SVR calibration

The second method we have used in our analysis is Support Vector Machines (SVM). SVM aims to classify the data into two groups by means of a hyperplane in a high-dimensional space. It can be written as a modified penalized logistic regression that solves the following objective function:

$$\hat{\beta}(C) = \underset{\beta \epsilon \mathbb{R}^k}{\mathrm{argmin}}\, C \left[ \sum_{i=1}^{n} y_i\, max\{0, 1 - K_i'\beta\} + (1 - y_i)\, max\{0, K_i'\beta - 1\} \right] + \|\beta\|_2$$

There are four terms here that differ from that in the standard logistic regression, and which allow SVM greater flexibility in covering a wide range of functional forms and dealing with high-dimensional data. In addition to the penalization factors, $C$ and $\|\beta\|_2$; there is a kernel $K$, and the *max(0,.)* component, which allows for some observations to be misclassified close to the separating margin. SVM has been found to achieve better performance than a number of other machine learning algorithms in some applications (Maroco et al., 2011). Nevertheless, given the similarity in terms of objective function, in other cases its performance has been found to be similar to that obtained with logistic regressions (Verplancke et al., 2008).

In practice, we have used an extension of SVM for continuous outcomes called Support Vector Regression (SVR). As SVM deviates for the logit model, SVR likewise deviates from OLS by using kernels to flexibly map the predictors into a hyperplane in a high-dimensional space, by allowing a margin in which it is insensitive to errors, and by adding a penalisation term (Guenther and Schonlau, 2016). One downside of SVM and SVR is that the parameters are difficult to interpret, and so it is not possible to say which variables matter most for determining the predictions made by this method.

We have implemented Support Vector Regression in Stata 16 using the command *svmachines* (Guenther and Schonlau 2016). A grid-search has been used to find the optimal parameters, as described in Appendix B.1. We have considered 5 possible values for each parameter (the penalization term $C$, the kernel smoothing parameter $\gamma$, and the bandwidth $\varepsilon$), as well as one kernel form (Guassian) for a total of 125 combinations ($1 \times 5^3$). As shown in Table B2, the selected values are in the interior of the provided intervals. Furthermore, as shown in Tables 2 and B4, the in-sample and out-of-sample MSE are close to each other, thus reducing any concern about overfitting, and confirming that the algorithms have been properly calibrated.

**Table B2: SVR tuning.**

| Outcome | Kernel | $C$ | | | | $\gamma$ | | | | $\varepsilon$ | | | |
| --- | --- | --- | --- | --- | --- | --- | --- | --- | --- | --- | --- | --- | --- |
| | | Min | Max | #val | Sel | Min | Max | #val | Sel | Min | Max | #val | Sel |
| Income support | Normal | 0.001 | 10 | 5 | 0.1 | 0.001 | 10 | 5 | 0.01 | 0.001 | 10 | 5 | 0.1 |

### B.4 Boosting calibration

The third method we have used in our analysis is Boosting (also called Boosted Regression), using the gradient boosting algorithm of Friedman et al. (2000). Gradient boosting is an ensemble method, which builds a classifier out of a large number of smaller classifiers. It works by combining many regression trees together. The basic idea behind regression trees is to split a sample sequentially based on a single covariate at a time exceeding a given threshold. The predicted outcome for each observation is then the average outcome in the final split. Regression trees are particularly effective when there is a large set of potential inputs, and only a few of them are actually strong predictors (Athey and Imbens, 2019).

As explained in Athey and Imbens (2019), boosting algorithms based on regression trees with one split can approximate any function that is additive in the features. Similarly, using regression trees with two splits allows for general second order effects, and so on. We consider trees with up to 6 splits, allowing up to 6-way interactions between variables. Thus, we model the function in a very flexible way. Boosting has been found to have superior performances than a number of other machine learning algorithms in many simulations (Bauer et al. 1999; Friedman et al. 2000) and has been used by Chalfin et al. (2016) in their work on predicting police hiring.

Boosting has been implemented in Stata using the command *boost* (Schonlau 2005). A grid-search has been used to find the optimal parameters, as described in Appendix B.1. In addition, we have used only a random subset of the train set (80% of it) to build the tree at each iteration. This technique - bagging - is used to reduce the variance of the final prediction without influencing the bias. We have also imposed a normal distribution for the error term and a shrinkage term (determining the contribution to each tree to the final selection) of 1. As shown in Table B3, we have considered trees with up to 6 tree splits, and up to 100 iterations (i.e., number of trees), for a total of 600 combinations (6*100). The selected values are in the interior of the provided intervals. Furthermore, as shown in Tables 2 and B4, the in-sample and out-of-sample MSE are close to each other, thus reducing any concern about overfitting, and confirming that the algorithms have been properly calibrated.

**Table B3: Boosting tuning.**

| Outcome | Distribution | Shrinkage | # tree splits | | | | # trees | | | |
|---|---|---|---|---|---|---|---|---|---|---|
| | | | Min | Max | #val | Sel | Min | Max | #val | Sel |
| Income support | Normal | 1 | 1 | 6 | 6 | 4 | 1 | 100 | 100 | 31 |

## B.5 Additional tables

## Table B4: Prediction (in-sample) accuracy for time on income support.

| | | Model | Predictors | MSE Train | MSE 5-Fold CV | $R^2$ |
|---|---|---|---|---|---|---|
| | 1 | OLS | Constant | 0.201 | - | 0.0% |
| Simple models | 2 | OLS | Sex | 0.199 | - | 0.9% |
| | 3 | OLS | Age | 0.165 | - | 17.8% |
| | 4 | OLS | Education | 0.193 | - | 3.5% |
| | 5 | OLS | Income support history | 0.082 | - | 59.3% |
| Economist models | 6 | OLS | Heuristic | 0.139 | - | 30.6% |
| | 7 | OLS | Heuristic + Income support history | 0.076 | - | 62.0% |
| | 8 | OLS | Model 7 + Location | 0.075 | - | 62.7% |
| Actuarial | 9 | OLS | Carer + Student + Parent | 0.188 | - | 6.4% |
| Machine Learning | 10 | LASSO | All baseline inputs | 0.053 | 0.055 | 73.8% |
| | 11 | SVR | All baseline inputs | 0.051 | 0.053 | 74.6% |
| | 12 | Boosting | All baseline inputs | 0.049 | 0.052 | 75.6% |
| | | | Full sample mean | 0.528 (N=50,615) | | |

*Notes*: This table report the in-sample performance of OLS and ML models. Each row is a different model. See also notes in Table 2.

## Table B5: Prediction (in-sample) accuracy for time on income support. Extensions.

| | Model | Predictors | MSE Train | MSE 5-Fold CV | $R^2$ |
|---|---|---|---|---|---|
| 1 | OLS | Top 10 predictors from the LASSO model | 0.076 | - | 62.3% |
| 2 | OLS | Top 10 predictors from the Boosting model | 0.060 | - | 70.0% |
| 3 | OLS | Union of the top predictors from Models 1 and 2 | 0.057 | - | 71.5% |
| 4 | OLS | Indigenous | 0.200 | - | 1.2% |
| 5 | OLS | Country of birth | 0.200 | - | 2.7% |
| 6 | OLS | Heuristic + Income support history + ZIP code | 0.070 | - | 65.0% |
| 7 | OLS | Heuristic + Parental welfare receipt | 0.080 | - | 33.2% |
| 8 | OLS | Heuristic + Income support history + Parental welfare receipt | 0.070 | - | 63.9% |
| 9 | FOR* | Heuristic | 0.140 | - | 31.2% |
| 10 | Boosting | Heuristic | 0.123 | 0.130 | 38.6% |
| 11 | LASSO | Heuristic | 0.139 | 0.140 | 30.6% |
| 12 | LASSO | All baseline inputs (plus interactions) | 0.050 | 0.052 | 75.3% |
| 13 | OLS | Heuristic (excluding those 100% on welfare 2011-14) | 0.120 | - | 32.8% |
| 14 | OLS | Heuristic + Income support history (excluding those 100% on welfare 2011-14) | 0.081 | - | 52.7% |
| 15 | LASSO | Full model (excluding those 100% on welfare 2011-14) | 0.058 | 0.059 | 67.0% |

*Notes*: This table report the in-sample performance of OLS and ML models. Each row is a different model. See also notes in Table 4.

**Table B6: Top ten predictors selected by the LASSO – Plus Interactions Model.**

| Order of Importance | Level of Importance | Predictors (calculated in 2014) | Variable Name |
|---|---|---|---|
| LASSO | Post-LASSO Coefficients | | |
| 1 | -0.972 | Benefit fluctuation | p_sdpy |
| 2 | 0.694 | Number of days received IS, Total benefit received | p_isdur14, p_totpy2014 |
| 3 | -0.488 | Receipt of Age Pension, Total benefit received | p_evage14, p_totpy2014 |
| 4 | -0.352 | Receipt of Disability Support Pension, Total benefit received | p_evdsp14, p_totpy2014 |
| 5 | 0.331 | Number of days received other types of IS payments, Total benefit received | p_ othdur2014, p_totpy2014 |
| 6 | 0.325 | Receipt of Age Pension, In Centrelink System in December | p_evage14, p_Dec2014 |
| 7 | 0.321 | Receipt of Income Support, In Centrelink System in December | p_evis14, p_Dec2014 |
| 8 | -0.315 | Total Employment Income | p_totinc2014 |
| 9 | 0.311 | Total benefit received, Ever acquired variable income spell missing | p_totpy2014, p_evacqvar2014miss |
| 10 | 0.294 | Receipt of Disability Support Pension, Received benefit in December | p_evdsp14, p_Dec2014 |

*Notes*: This variable list the top 10 inputs selected by LASSO (Model 15 Table 4) when predicting the proportion of time an individual received an income support payment in 2015-2019. All listed variables were measured in 2014. We manually coded two-way interaction variables based on the top 20 predictors from the base LASSO algorithm (Model 10, Table 2). LASSO (interaction model) selected 495 variables. In rows where there are two variables listed as opposed to one variable, LASSO has selected the interaction term. In LASSO, the reported coefficients are computed by taking the selected variables and estimating an OLS linear regression with them. All base variables are described in Online Appendix A.

**Table B7: Prediction (in-sample) accuracy for time on Unemployment Benefits.**

|  |  | Model | Predictors | MSE Mean | MSE 5-Fold CV | $R^2$ |
|---|---|---|---|---|---|---|
| Simple models | 1 | OLS | Constant | 0.206 | - | 0.0% |
|  | 2 | OLS | Sex | 0.204 | - | 1.0% |
|  | 3 | OLS | Education | 0.199 | - | 3.3% |
|  | 4 | OLS | Age | 0.086 | - | 58.3% |
|  | 5 | OLS | Income support history | 0.126 | - | 39.2% |
| Economist models | 6 | OLS | Heuristic | 0.036 | - | 82.8% |
|  | 7 | OLS | Heuristic + Income support history | 0.035 | - | 83.1% |
|  | 8 | OLS | Model 7 + Location | 0.035 | - | 83.5% |
| Machine Learning | 9 | LASSO | All baseline inputs | 0.029 | 0.032 | 86.2% |

*Notes*: The outcome variable is the proportion of time an individual received an unemployment benefit in 2015-2019. Each row is a different model. See also notes in Table 5. All variables are described in Online Appendix A.

**Table B8: Clustering: Key characteristics.**

| Predictors | Group 1 | Group 2 | Group 3 | Group 4 | No benefit |
|---|---|---|---|---|---|
| Government benefit fluctuation | 0.012 | 0.011 | 0.012 | 0.016 | 0.016 |
|  | (0.011) | (0.01) | (0.009) | (0.011) | (0.015) |
| Total amount of benefit received | 0.219 | 0.198 | 0.142 | 0.271 | 0.05 |
|  | (0.09) | (0.1) | (0.08) | (0.127) | (0.045) |
| Government benefit durations | 0.967 | 0.937 | 0.894 | 0.943 | 0.021 |
|  | (0.14) | (0.186) | (0.225) | (0.177) | (0.097) |
| Annual employment income | 0.009 | 0.015 | 0.019 | 0.015 | 0.004 |
|  | (0.028) | (0.046) | (0.04) | (0.042) | (0.033) |
| Wage rate | 0.000 | 0.000 | 0.000 | 0.000 | 0.000 |
|  | (0.001) | (0.001) | (0.000) | (0.002) | (0.000) |
| Number of residential moves | 0.009 | 0.009 | 0.021 | 0.013 | 0.006 |
|  | (0.019) | (0.022) | (0.032) | (0.021) | (0.013) |
| Number of sanctions | 0.005 | 0.005 | 0.009 | 0.003 | 0.002 |
|  | (0.023) | (0.026) | (0.031) | (0.018) | (0.014) |
| Age – 15 years old (at 1Jan2014) | 0.004 | 0.001 | 0.009 | 0.003 | 0.000 |
|  | (0.062) | (0.036) | (0.094) | (0.054) | (0.000) |
| Simultaneous jobs worked (num) | 0.013 | 0.013 | 0.025 | 0.015 | 0.004 |
|  | (0.029) | (0.03) | (0.037) | (0.039) | (0.022) |
| Payment receipt in December | 0.996 | 0.99 | 0.991 | 0.992 | 0.162 |
|  | (0.062) | (0.098) | (0.094) | (0.089) | (0.368) |
| Female | 0.419 | 0.574 | 0.468 | 0.704 | 0.657 |
|  | (0.494) | (0.495) | (0.499) | (0.457) | (0.475) |
| Immigrant | 0.252 | 0.314 | 0.194 | 0.200 | 0.368 |
|  | (0.435) | (0.464) | (0.396) | (0.400) | (0.483) |
| Immigrant missing indicator | 0.004 | 0.000 | 0.000 | 0.001 | 0.001 |
|  | (0.062) | (0.018) | (0.000) | (0.031) | (0.037) |
| Indigenous | 0.081 | 0.051 | 0.155 | 0.092 | 0.013 |
|  | (0.274) | (0.219) | (0.362) | (0.289) | (0.112) |
| Indigenous missing indicator | 0.066 | 0.074 | 0.07 | 0.047 | 0.035 |
|  | (0.249) | (0.262) | (0.256) | (0.211) | (0.184) |
| Aged below 25 | 0.116 | 0.037 | 0.355 | 0.332 | 0.070 |
|  | (0.321) | (0.19) | (0.479) | (0.471) | (0.256) |
| Aged 25-29 | 0.070 | 0.036 | 0.116 | 0.166 | 0.135 |
|  | (0.255) | (0.186) | (0.321) | (0.372) | (0.342) |
| Aged 30-34 | 0.085 | 0.042 | 0.093 | 0.160 | 0.256 |
|  | (0.28) | (0.201) | (0.291) | (0.366) | (0.437) |
| Aged 35-39 | 0.093 | 0.081 | 0.067 | 0.075 | 0.226 |
|  | (0.291) | (0.273) | (0.249) | (0.264) | (0.419) |
| Aged 40-44 | 0.081 | 0.095 | 0.069 | 0.058 | 0.154 |
|  | (0.274) | (0.294) | (0.254) | (0.235) | (0.362) |
| Aged 45-49 | 0.116 | 0.099 | 0.075 | 0.045 | 0.100 |
|  | (0.321) | (0.298) | (0.263) | (0.207) | (0.300) |
| Aged 50-54 | 0.120 | 0.106 | 0.086 | 0.043 | 0.046 |
|  | (0.326) | (0.309) | (0.281) | (0.202) | (0.21) |
| Aged 55-59 | 0.081 | 0.122 | 0.060 | 0.049 | 0.010 |
|  | (0.274) | (0.327) | (0.238) | (0.215) | (0.099) |
| Aged 60-64 | 0.105 | 0.211 | 0.078 | 0.044 | 0.003 |
|  | (0.307) | (0.408) | (0.269) | (0.204) | (0.053) |
| Year 10 or below | 0.000 | 0.000 | 0.285 | 0.434 | 0.028 |
|  | (0.000) | (0.000) | (0.452) | (0.496) | (0.165) |
| Year 12 or below | 0.004 | 0.000 | 0.579 | 0.870 | 0.146 |
|  | (0.062) | (0.000) | (0.494) | (0.336) | (0.353) |
| Certificate or below | 0.008 | 0.000 | 0.596 | 0.900 | 0.156 |
|  | (0.088) | (0.000) | (0.491) | (0.300) | (0.363) |
| Certificate 1 or below | 0.109 | 0.000 | 0.668 | 0.912 | 0.157 |
|  | (0.312) | (0.000) | (0.471) | (0.284) | (0.364) |
| Certificate 2 or below | 0.512 | 0.000 | 0.973 | 0.993 | 0.166 |
|  | (0.501) | (0.000) | (0.161) | (0.083) | (0.372) |
| Certificate 3 or below | 0.527 | 0.001 | 0.997 | 0.997 | 0.167 |
|  | (0.500) | (0.025) | (0.052) | (0.054) | (0.373) |
| Certificate 4 or below | 0.713 | 0.023 | 0.999 | 1.000 | 0.173 |
|  | (0.453) | (0.148) | (0.03) | (0.000) | (0.378) |
| Diploma or below | 0.895 | 0.056 | 1.000 | 1.000 | 0.181 |
|  | (0.307) | (0.229) | (0.000) | (0.000) | (0.385) |
| Bachelor's degree or below | 0.996 | 0.087 | 1.000 | 1.000 | 0.247 |
|  | (0.062) | (0.282) | (0.000) | (0.000) | (0.432) |
| Education missing | 0.000 | 0.905 | 0.000 | 0.000 | 0.744 |
|  | (0.000) | (0.293) | (0.000) | (0.000) | (0.437) |
| Ever married in 2014 | 0.221 | 0.356 | 0.143 | 0.138 | 0.736 |
|  | (0.416) | (0.479) | (0.35) | (0.345) | (0.441) |
| Ever married missing indicator | 0.000 | 0.000 | 0.002 | 0.000 | 0.000 |
|  | (0.000) | (0.000) | (0.042) | (0.000) | (0.000) |
| With child in 2014 | 0.240 | 0.308 | 0.277 | 0.548 | 0.751 |
|  | (0.428) | (0.462) | (0.448) | (0.498) | (0.433) |
| New South Wales | 0.349 | 0.335 | 0.307 | 0.321 | 0.351 |
|  | (0.478) | (0.472) | (0.462) | (0.467) | (0.478) |
| Victoria | 0.256 | 0.249 | 0.213 | 0.250 | 0.261 |
|  | (0.437) | (0.432) | (0.41) | (0.433) | (0.44) |
| Queensland | 0.225 | 0.197 | 0.266 | 0.250 | 0.192 |
|  | (0.418) | (0.397) | (0.442) | (0.433) | (0.394) |
| Western Australia | 0.085 | 0.089 | 0.099 | 0.082 | 0.090 |
|  | (0.28) | (0.285) | (0.299) | (0.275) | (0.286) |
| South Australia | 0.085 | 0.091 | 0.111 | 0.077 | 0.083 |
|  | (0.28) | (0.287) | (0.314) | (0.267) | (0.276) |
| Australian Capital Territory | 0.000 | 0.012 | 0.012 | 0.010 | 0.018 |
|  | (0.000) | (0.107) | (0.107) | (0.099) | (0.134) |
| Northern Territory | 0.004 | 0.013 | 0.021 | 0.014 | 0.022 |
|  | (0.062) | (0.114) | (0.144) | (0.117) | (0.148) |
| Tasmania | 0.019 | 0.037 | 0.031 | 0.039 | 0.008 |
|  | (0.138) | (0.188) | (0.174) | (0.193) | (0.091) |
| Any type of income support | 1.000 | 1.000 | 1.000 | 1.000 | 0.066 |
|  | (0.000) | (0.000) | (0.000) | (0.000) | (0.248) |
| Unemployment benefits | 0.012 | 0.22 | 0.993 | 0.001 | 0.038 |
|  | (0.107) | (0.415) | (0.084) | (0.031) | (0.191) |
| Disability Support Pension | 0.566 | 0.402 | 0.017 | 0.321 | 0.000 |
|  | (0.497) | (0.49) | (0.129) | (0.467) | (0.000) |
| Carer Payment | 0.194 | 0.108 | 0.029 | 0.104 | 0.000 |
|  | (0.396) | (0.31) | (0.169) | (0.305) | (0.000) |
| Parenting payments | 0.004 | 0.077 | 0.084 | 0.424 | 0.000 |
|  | (0.062) | (0.266) | (0.278) | (0.494) | (0.000) |
| Age Pension | 0.147 | 0.217 | 0.012 | 0.033 | 0.000 |
|  | (0.355) | (0.412) | (0.111) | (0.178) | (0.000) |
| Partner payments | 0.000 | 0.014 | 0.000 | 0.001 | 0.000 |
|  | (0.000) | (0.118) | (0.000) | (0.031) | (0.000) |
| Observations | 258 | 3108 | 1126 | 1009 | 712 |

*Notes*: Summary statistics (mean and standard deviation) are reported for each group identified by the hierarchical clustering algorithm. Individuals who were identified as at-risk of long-term welfare receipt have been divided into four groups (first four column of results). The last column reports summary statistics for the group of individuals predicted to have no future receipt. All variables have been rescaled to be between 0 and 1. All variables are described in Online Appendix A.